\documentclass[10pt,journal,compsoc]{IEEEtran}
\usepackage{url}

\usepackage[normalem]{ulem}

\usepackage{fancyhdr}
\usepackage[normalem]{ulem}

\usepackage{enumitem}

\usepackage{array}

\usepackage{listings}
\usepackage{xcolor}

\definecolor{darkgreen}{rgb}{0,0.5,0}
\definecolor{tinygray}{rgb}{0.9,0.9,0.9}

\DeclareMathAlphabet{\mathcal}{OMS}{cmsy}{m}{n}

\usepackage[noend]{algpseudocode}
\usepackage{algorithm}

\usepackage{textcase}

\usepackage{tikz}

\usepackage{multirow}

\usepackage[skip=5pt]{caption}
\usepackage{listings}
\usepackage{lipsum}

\usepackage{mathtools}

\ifCLASSOPTIONcompsoc
  \usepackage[nocompress]{cite}
\else
  \usepackage{cite}
\fi
\ifCLASSINFOpdf
\else
\fi
\begin{document}
%
\title{Evaluating Modern GPU Interconnect: PCIe, NVLink, NV-SLI, NVSwitch and GPUDirect}

\author{Ang~Li,~Shuaiwen~Leon~Song,~Jieyang~Chen,~Jiajia~Li,~Xu~Liu,~Nathan~Tallent, and~Kevin~Barker
\IEEEcompsocitemizethanks{\IEEEcompsocthanksitem Ang Li is a research scientist in Pacific Northwest National Laboratory.\protect\\
E-mail: ang.li@pnnl.gov}
\thanks{Manuscript received January 19, 2019.}}

\IEEEtitleabstractindextext{%
\begin{abstract}
High performance multi-GPU computing becomes an inevitable trend due to the ever-increasing demand on computation capability in emerging domains such as deep learning, big data and planet-scale simulations. However, the lack of deep understanding on how modern GPUs can be connected and the real impact of state-of-the-art interconnect technology on multi-GPU application performance become a hurdle. In this paper, we fill the gap by conducting a thorough evaluation on five latest types of modern GPU interconnects: PCIe, NVLink-V1, NVLink-V2, NVLink-SLI and NVSwitch, from six high-end servers and HPC platforms: NVIDIA P100-DGX-1, V100-DGX-1, DGX-2, OLCF's SummitDev and Summit supercomputers, as well as an SLI-linked system with two NVIDIA Turing RTX-2080 GPUs. Based on the empirical evaluation, we have observed four new types of GPU communication network NUMA effects: three are triggered by NVLink's topology, connectivity and routing, while one is caused by PCIe chipset design issue. These observations indicate that, for an application running in a multi-GPU node, choosing the right GPU combination can impose considerable impact on GPU communication efficiency, as well as the application's overall performance. Our evaluation can be leveraged in building practical multi-GPU performance models, which are vital for GPU task allocation, scheduling and migration in a shared environment (e.g., AI cloud and HPC centers), as well as communication-oriented performance tuning.
\end{abstract}

}

\maketitle

\IEEEdisplaynontitleabstractindextext

%
\IEEEpeerreviewmaketitle

\IEEEraisesectionheading{\section{Introduction}\label{sec:introduction}}
\IEEEPARstart{M}{ulti-GPU} execution nowadays becomes an inevitable trend for warehouse GPGPU computing. This is due to the ever-increasing demand of computation capability from emerging domains such as machine learning, big data and planet-scale simulations \cite{goyal2017accurate, fuhrernear}. With increasingly larger problem to solve, scalable GPU computing becomes necessary. Recently, a research group from \emph{Sony} leveraged 3,456 GPUs to train a \emph{ResNet-50} neural network for \emph{ImageNet} in 122 seconds, achieving near optimal GPU scaling efficiency \cite{mikami2018imagenet}. The \emph{Swiss National Supercomputing Center} (\emph{CSCS}) relied on 4,888 GPUs in the \emph{Piz Daint} supercomputer to simulate near-global climate in ultra-high resolution \cite{fuhrernear}. 

Multi-GPU execution scales in two directions: \emph{vertically scaling-up in a single node} and \emph{horizontally scaling-out across multiple nodes}. Good examples to describe the intra-node scale-up scenario are the latest NVIDIA DGX-1 \cite{nvidia2017dgx1} and DGX-2 \cite{nvidia2018dgx2} super-AI servers, which incorporate 8 and 16 P100/V100 GPUs connected by NVLink and NVSwitch, respectively. For the inter-node scale-out scenario, the U.S. \emph{Department of Energy} (\emph{DOE}) has recently deployed two GPU-accelerated supercomputers \emph{Summit} \cite{summit} and \emph{Sierra} \cite{sierra} in \emph{Oak Ridge} and \emph{Livermore National Laboratories}, with more than 3400 GPU-integrated nodes interconnected. 



Gaining performance from multi-GPU scaling, however, is not trivial, mainly because (i) There are no mature multi-GPU parallel programming, execution and performance models, largely due to the limited knowledge on how modern GPUs are interconnected as well as their communication patterns; (ii) Traditionally, inter-GPU communication shares the same bus interconnect as CPU-GPU communication, such as PCIe. This  situation recently changed due to the introduction of GPU-oriented interconnect such as NVLink, NV-SLI and NVSwitch. However, their characteristics, as well as the performance impact on real-world multi-GPU applications are still unknown, limiting the efforts to leverage them for advanced performance tuning and delivery.

In this paper, we fill this gap by thoroughly characterizing a variety of modern GPU interconnects, including PCIe, NVLink Version-1, NVLink Version-2, NV-SLI, NVSwitch, and GPUDirect. We measured their raw startup latency, sustainable uni/bi-directional bandwidth, network topology, communication efficiency, routing, and NUMA effects, under the two communication patterns: \emph{Peer-to-Peer (P2P)} and \emph{Collective (CL)}. Based on these results, we summarize several observations, challenges to address, and potential research topics regarding to multi-GPU execution. Through this evaluation, software designers can gain deeper knowledge about the latest GPU interconnects, paving the way for building more mature multi-GPU programming, execution and performance models, and reliable simulators for better guiding application development and performance tuning.




\section{Modern GPU Interconnect}
\label{sec_interconnect}

We focus on six types of modern GPU interconnect: \emph{PCIe}, \emph{NVLink-V1}, \emph{NVLink-V2}, \emph{NV-SLI}, \emph{NVSwitch}, and \emph{GPUDirect-enabled InfiniBand}. Table~\ref{tab:platform} lists the platforms we used for evaluation. For GPU-GPU communication, \emph{P100-DGX-1}, \emph{V100-DGX-1} are for evaluating PCIe, NVLink-V1 and NVLink-V2. \emph{SLI} is for NV-SLI. \emph{DGX-2} is for \emph{NVSwitch}. \emph{SummitDev} and \emph{Summit} are for assessing inter-node InfiniBand interconnect with GPUDirect-RDMA. We first briefly review every technology.

\begin{table*}[!t]
\centering\scriptsize
\caption{Evaluation Platforms. ``Arch'' refers to GPU architecture generation. ``SP/DP GFlops'' refer to GPU theoretical single/double floating-point performance. ``Rtm'' refers to CUDA runtime version.}
\begin{tabular}{|c|l|c|c|c|c|c|c|c|c|c|}
\hline
\textbf{Platform} & \textbf{Configuration} & \textbf{Interconnect}  & \textbf{CPU} & \textbf{Compiler} & \textbf{GPU}  & \textbf{GPU Arch} & \textbf{SP/DP GFlops} & \textbf{GPU Memory} & \textbf{Rtm} \\ \hline
P100-DGX-1 & Single node, 8 GPUs & NVLink-V1 & Intel Xeon E5-2698  & gcc-4.8.4 & Tesla-P100 & Pascal & 10609/5304 & 16GB HBM2 @ 732 GB/s & 8.0  \\ \hline
V100-DGX-1 & Single node, 8 GPUs & NVLink-V2 & Intel Xeon E5-2698  & gcc-5.4.0 & Tesla-V100 & Volta & 14899/7450 & 16GB HBM2 @ 900 GB/s & 9.0 \\ \hline
DGX-2 & Single node, 16 GPUs & NVSwitch & Intel Xeon P-8168  & gcc-7.3.0 & Tesla-V100 & Volta & 14899/7450 & 16GB HBM2 @ 900 GB/s & 9.0 \\ \hline
SLI & Single node, 2 GPUs & NVLink-SLI & Intel Xeon E5-2680 & gcc-4.8.5 & RTX-2080 & Turing & 10068/314.6  & 8GB GDDR6 @ 448 GB/s &  10.0  \\ \hline
SummitDev & 54 nodes, 4 GPUs/node & NVLink-V1 & IBM Power-8  & xlc-13.1.6 & Tesla-P100 & Pascal & 10609/5304 & 16GB HBM2 @ 732 GB/s & 8.0  \\ \hline
Summit & 4600 nodes, 6 GPUs/node & NVLink-V2 & IBM Power-9  & xlc-16.1.1 & Tesla-V100 & Volta & 14899/7450 & 16GB HBM2 @ 900 GB/s & 9.2  \\ \hline
\end{tabular}
\label{tab:platform}
\end{table*}

\subsection{PCIe}
\emph{Peripheral-Component-Interconnect-Express-Bus} (PCIe), is a high-speed serial computer expansion bus standard. Traditionally, a GPU-integrated system connect one or multiple GPU devices to the CPUs via PCIe. However, compared to the interconnect between CPU and DRAM, PCIe is much slower. It often becomes a major performance bottleneck for GPU-acceleration \cite{pabst2010fast, xu2014graph, pcie}. Such a condition exacerbates when PCIe based GPU P2P communication is adopted \cite{nvidia2017dgx1}.

\begin{figure}[!t]
\centering
\includegraphics[width=0.95\columnwidth]{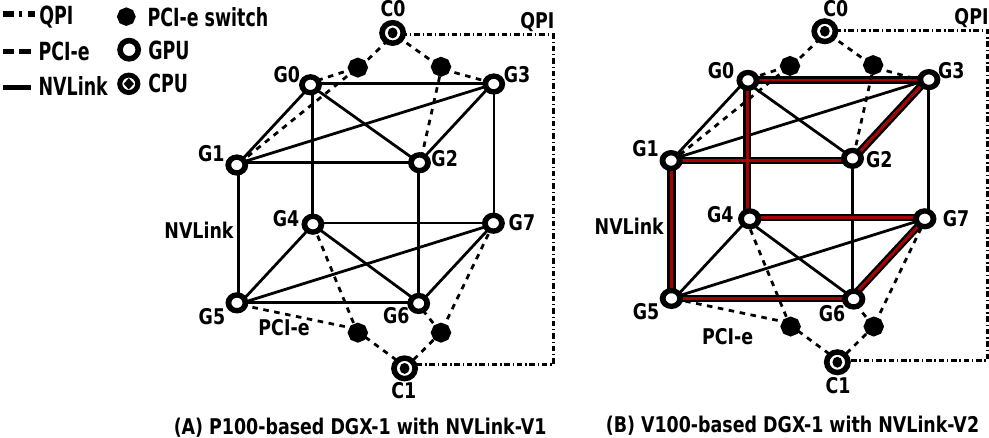} 
\caption{PCIe and NVLink-V1/V2 topology for P100-DGX-1 and V100-DGX-1.}
\label{fig:dgx-1}
\end{figure}

The dash-lines in Figure~\ref{fig:dgx-1}-(A) illustrate how the eight GPUs are interconnected by PCIe (and QPI) in DGX-1. As is shown, the PCIe network in DGX-1 forms a balanced tree structure, e.g., GPU-0 and GPU-1 are connected via a PCIe switch. The switch is further connected to CPU Socket-0. Similar scenarios apply to other GPUs. Finally, the dual CPU sockets are bridged by \emph{QuickPath Interconnect} or QPI \cite{ziakas2010intel}. PCIe in DGX-2 also form a tree-based topology, but adopts two-level PCIe switches, as shown in Figure~\ref{fig:nvswitch-pcie}.

\subsection{NVLink-V1}

Known as the first generation of NVLink, NVLink-V1 is a wire-based communication interface for near-range devices based on \emph{High-Speed-Signaling-Interconnect} (NVHS) \cite{nvidia2017dgx1, foley2017ultra}. It supports P2P communication that enables CPU-GPU or GPU-GPU linking. It allows direct read and write on remote CPU's host-memory and/or peer GPU's device-memory. Remote atomic operations are also feasible. NVLink is bidirectional; each link consists of two sublinks --- one for each direction. Each sublink further contains eight differential NVHS lanes. An embedded clock is integrated for transmission.  The packet size varies from 16 bytes to 256 bytes (one 128-bit flit to sixteen 128-bit flits). The communication efficiency is strongly correlated to the packet size. Overall, it is reported to be twice as efficient as PCIe \cite{foley2017ultra}.


An NVLink can be viewed as a cable with two terminal-plugs whereas each GPU incorporates several NVLink slots. How these slots are connected via the NVLink cables dictate the topology and bandwidth of the GPU network. Multiple cables can be ganged together to enhance bandwidth when they are linking the same endpoints. A Pascal-P100 GPU has quad NVLink slots. Therefore, for a dual-GPU system, a direct setting would be two GPUs connected by four NVLinks, leading to 4$\times$ bandwidth of a single link.


\vspace{4pt}\noindent\emph{P100-DGX-1:} The GPU network topology for DGX-1 is known as \emph{Hypercube Mesh}. As shown in Figure~\ref{fig:dgx-1}-(A), each GPU occupies a corner of the cube and all the 12 edges are NVLink connections. For the upper and lower planes, the diagonals are also connected, forming two fully-connected subsets. Such a topology design is balanced, with stronger connectivity inside a plane. In other words, accessing within a plane is UMA, while accessing nodes across planes leads to NUMA (when they are not directly linked, e.g., from GPU-0 to GPU-7). In fact, NVLink is not self-routed when the two terminals are not directly linked. It relies on explicit routing through a user-specified intermediate node.

\begin{figure}[!t]
\centering
\includegraphics[width=0.8\columnwidth]{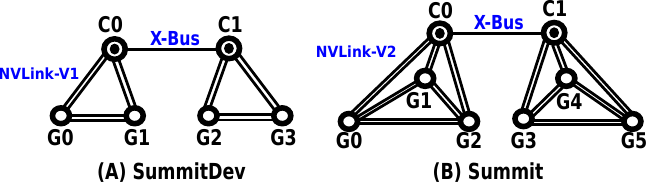} 
\caption{NVLink interconnect topology for SummitDev and Summit.}
\label{fig:summit}
\end{figure}

\vspace{4pt}\noindent\emph{SummitDev:} The interconnect topology inside a machine node in SummitDev is illustrated in Figure~\ref{fig:summit}-(A). As can be seen, the four P100 GPUs per node are partitioned into two subsets; two GPUs per subset. A subset, together with a Power-8 CPU socket, constituting a triple-node fully-connected subnetwork. Every two nodes (either CPU or GPU) in a subnetwork are connected by two NVLink-V1 links. The two subnetworks communicate via an X-Bus at 38 GB/s. Note, unlike DGX-1, there is no direct connection between GPUs from separated subnetworks, as all the four NVLink slots of the P100 GPUs have already been occupied. 

\subsection{NVLink-V2}

\begin{figure*}[!htb]
\minipage{0.99\columnwidth}
\centering
\includegraphics[width=0.95\columnwidth]{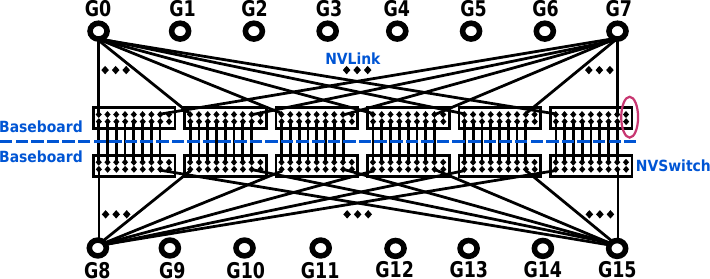} 
\caption{NVSwitch interconnect topology in DGX-2.}
\label{fig:nvswitch}
\endminipage\hfill
\minipage{0.99\columnwidth}
\centering
\includegraphics[width=0.9\columnwidth]{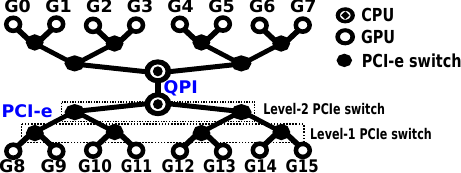} 
\caption{PCIe interconnect topology in DGX-2.}
\label{fig:nvswitch-pcie}
\endminipage
\end{figure*}

The second generation of NVLink improves per-link bandwidth and adds more link-slots per GPU: in addition to 4 link-slots in P100, each V100 GPU features 6 NVLink slots; the bandwidth of each link is also enhanced by 25\%. Besides, a low-power operating mode is introduced for saving power in case a link is not being heavily exploited. The extra two link-slots have enabled certain alternation to the original network topology. 

\vspace{4pt}\noindent\emph{V100-DGX-1:}  Shown in Figure~\ref{fig:dgx-1}-(B), the V100-DGX-1 does not choose to further strengthen the connectivity within a plane, but forming a fast  \emph{Backbone Ring} inside the Hypercube-Mesh network. Each connection in this ring enables 2$\times$ bandwidth compared to other links. We suspect this is to improve the efficiency of collective communication, as further discussed in Section~\ref{subsec_intracl}.

\vspace{4pt}\noindent\emph{Summit:} Figure~\ref{fig:summit}-(B) shows the interconnect network topology for a machine node in Summit. The six GPUs are also organized in two subnetworks, but each with three GPUs. A subset, together with a Power-9 CPU socket, forming a quad-node fully-connect subnetwork. Every two nodes (either CPU or GPU) in a subnetwork are connected by two NVLink-V2 links. The two subnetworks are connected again via an X-Bus, at 64 GB/s. With one more node in the subnetwork, all the 6 NVLink slots per V100 GPU are fully utilized; there is no GPU direct connection between subsets.

\subsection{NVLink-SLI}

\emph{Scalable Link Interface} (SLI) \cite{nvidia2007sli}, or \emph{Crossfire} \cite{ati2009cross}, are traditionally used for graphic purposes only \cite{kim2014multi}. However, the recent Turing architecture GPUs (e.g., TU102, TU104) bring with them a new form of high-speed multi-GPU bridge, based on the NVLink-V2 interconnect technology. The bridge pairs up two GPUs so they can communicate with each other and co-render games, co-run GPGPU tasks, or co-share GPU memory spaces. In our SLI platform, the TU104 based RTX2080 GPU offers one x8 NVLink-V2 links, with up to 25 GB/s per direction per link, delivering an aggregate bidirectional bandwidth of 50 GB/s. Only two-way NVLink-SLI is supported for Turing GPUs.

\subsection{NVSwitch}

\emph{NVSwitch} \cite{nvswitch} is proposed mainly to address all-to-all communication in many emerging applications such as deep neural network training. NVSwitch currently only appears in DGX-2. The topology for NVSwitch and PCIe in DGX-2 are shown in Figure~\ref{fig:nvswitch} and Figure~\ref{fig:nvswitch-pcie}, respectively. NVSwitch is an NVLink-V2 based switch chip for intra-node communication, featuring 18 ports of NVLink per switch. Shown in Figure~\ref{fig:nvswitch}, there are two baseboards; each contains 6 NVSwitches and hosts 8 GPUs. This is because a V100 GPU incorporates 6 NVLink slots, being able to connect to 6 NVSwitches simultaneously, each target per link, at 50 GB/s bidirectional bandwidth. Each NVSwitch is a 18x18 fully connected non-blocking crossbar: (1) 8 of the 18 ports are used for intra-baseboard communication, which means any of the 8 GPUs on one baseboard can talk with any other GPUs on the same baseboard at a full bandwidth of 50 GB/s $\times$ 6 switches =300 GB/s via a single NVSwitch hop; (2) Another 8 of the 18 ports are are used to connect to the opposite baseboard, meaning that each of the GPUs on one baseboard can talk to any GPUs on the other baseboard also at a full bandwidth of 300 GB/s, but through 2 NVSwitch hops. The baseboard-to-baseboard raw bisection bandwidth is thus 25 GB/s $\times$ 8 links/switch $\times$ 6 switches=2.4 TB/s; (3) The left 2 ports per NVSwitch are reserved (e.g., the red circle in Figure~\ref{fig:nvswitch}).

To keep data integrity, \emph{Cyclical Redundancy Coding} (CRC) is imposed to detect NVLink transfer errors and replay the transfer when necessary. \emph{Error-Correcting Codes} (ECC) is enabled for the datapaths, routing, and state structures. In addition, final hop-address fidelity checks and buffer over/under-flow checks can be enforced in NVSwitch. To avoid illegal out-of-range access, a fabric manager monitors the routing tables for each particular application.

\subsection{InfiniBand with GPUDirect-RDMA}

\emph{GPUDirect InfiniBand}: We will not discuss InfiniBand \cite{grun2010introduction} itself since it is already widely used for HPC platforms today and has been extensively studied. Instead, we focus on its relation with GPU. Since the Kepler architecture, NVIDIA GPUs have introduced \textit{GPUDirect-RDMA} \cite{gpudirectrdma} (Correspondingly, AMD proposed \emph{ROCn-RDMA} \cite{rocnrdma}). It enables third-party PCIe devices, especially the IB \emph{Host-Channel-Adapter} (i.e., HCA) to directly access GPU device memory via PCIe without any assistance from CPU or staging through the main memory, which significantly improves the efficiency of inter-node GPU communication. To achieve IB RDMA, GPU vendors offer an OS kernel extension to return a DMA bus mapping for GPU device memory. When a user creates an IB region, it signals the IB driver with the target address of the GPU device memory. IB driver then calls a routine to obtain the DMA mapping. Finally, a normal IB virtual memory structure is returned to the user program as if it targets normal CPU memory. GPUDirect-RDMA is enabled for both SummitDev and Summit.


%

\begin{table*}[!t]
\centering\scriptsize
\caption{Tartan Microbenchmarks.}
\begin{tabular}{|c|c|c|c|l|}
\hline
\textbf{Scaling} & \textbf{Communication} & \textbf{Interconnect} & \textbf{Metrics}  & \textbf{Description}  \\ \hline
Scale-up & Peer-to-Peer & PCIe, NVLink-V1 and V2 &  Latency, bandwidth and efficiency &  Developed based on   \emph{p2pBandwidthLatencyTest} from CUDA SDK \cite{nvidia2015sdk} \\ \hline
Scale-up & Collective & PCIe, NVLink-V1 and V2 & Latency, bandwidth and efficiency &  Developed based on \emph{nccl-tests} \cite{nccl_tests} linking to NCCL-V1 \cite{ncclV1} and V2 \cite{ncclV2}  \\ \hline
Scale-out & Peer-to-Peer & InfiniBand with GPUDirect & Latency, bandwidth and efficiency &  Developed based on \emph{MPI-GPU-BW} \cite{mpi-gpu-bw} \\ \hline
Scale-out & Collective & InfiniBand with GPUDirect & Latency, bandwidth and efficiency &  Developed based on \emph{nccl-tests} \cite{nccl_tests} linking to NCCL-V2 \cite{ncclV2}  \\ \hline
\end{tabular}
\label{tab:micro}
\end{table*}

\begin{figure*}[!htb]
\minipage{0.65\columnwidth}
\includegraphics[width=1.05\columnwidth]{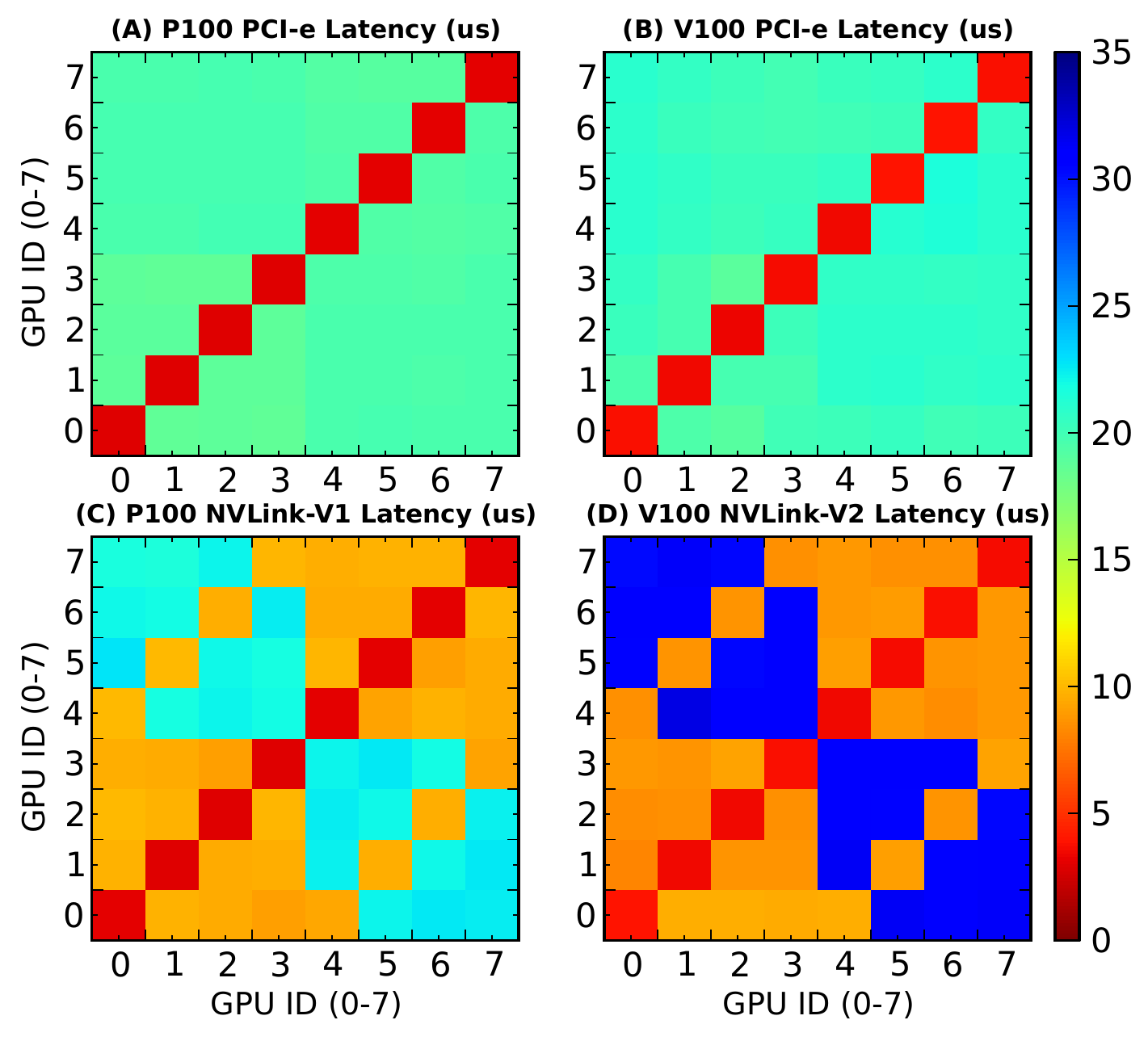} 
\caption{P100/V100-DGX-1 P2P communication latency. The red blocks along the anti-diagonal are local communication. The fact that other blocks are all green in (A) and (B) indicate that no NUMA effect appears on PCIe for latency. The orange and blue blocks in (C) and (D) are refer to neighbor nodes and remote nodes respectively which exhibit clear disparity.}
\label{fig:latency}
\endminipage\hfill
\minipage{0.66\columnwidth}
\includegraphics[width=1.05\columnwidth]{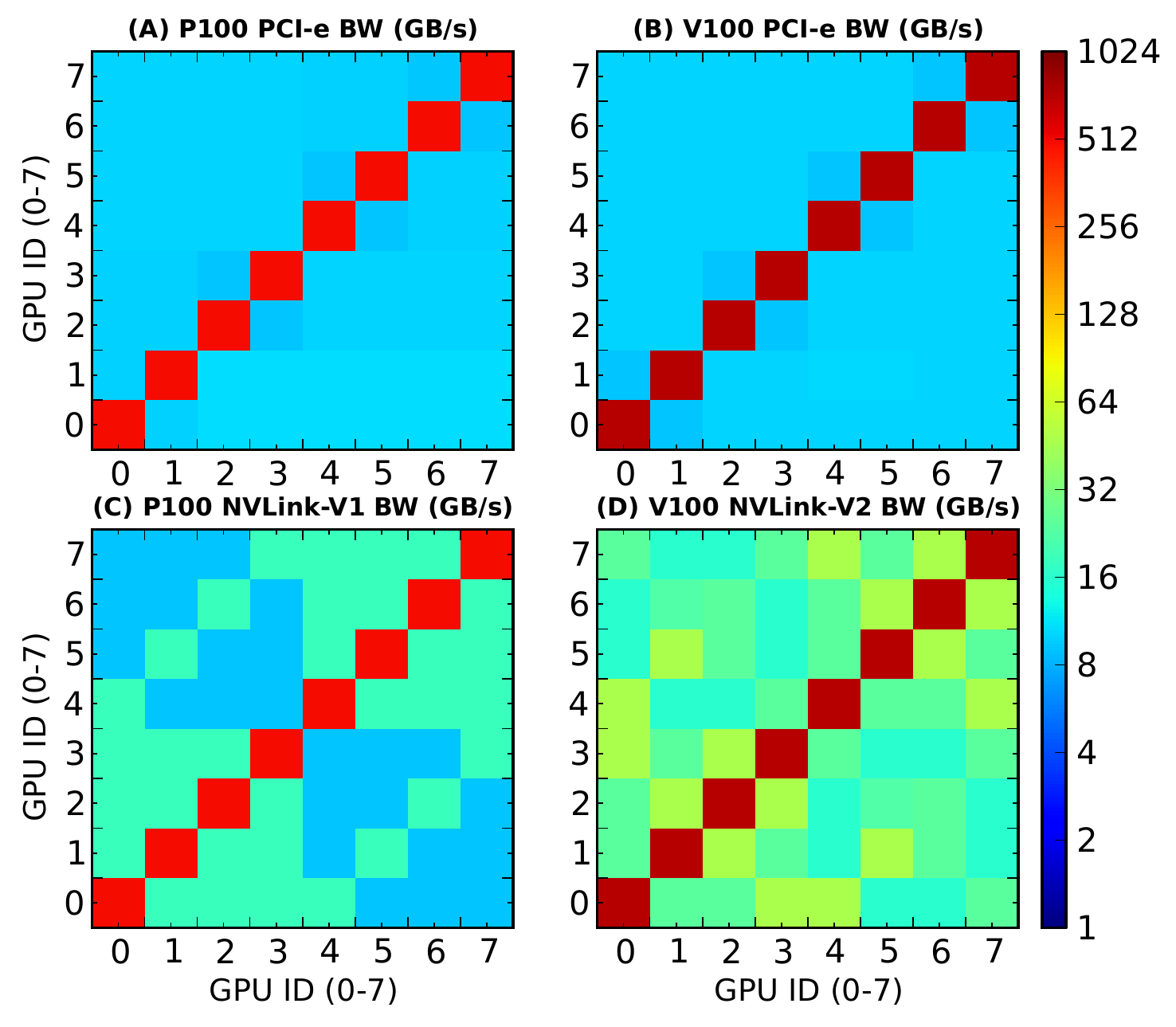} 
\caption{DGX-1 P2P unidirectional bandwidth. Although not very obvious, we can see 2x2 blocks in (A) and (B) along anti-diagonal, which indicates the anti-locality NUMA effect for unidirection bandwidth on PCIe. (C) and (D) confirm NUMA among neighbors and remote nodes. The other two types of NUMA for NVLink-V2 are not quite clear in (D). They are more obvious for bidirection bandwidth.}
\label{fig:unibandwidth}
\endminipage\hfill
\minipage{0.66\columnwidth}
\includegraphics[width=1.05\columnwidth]{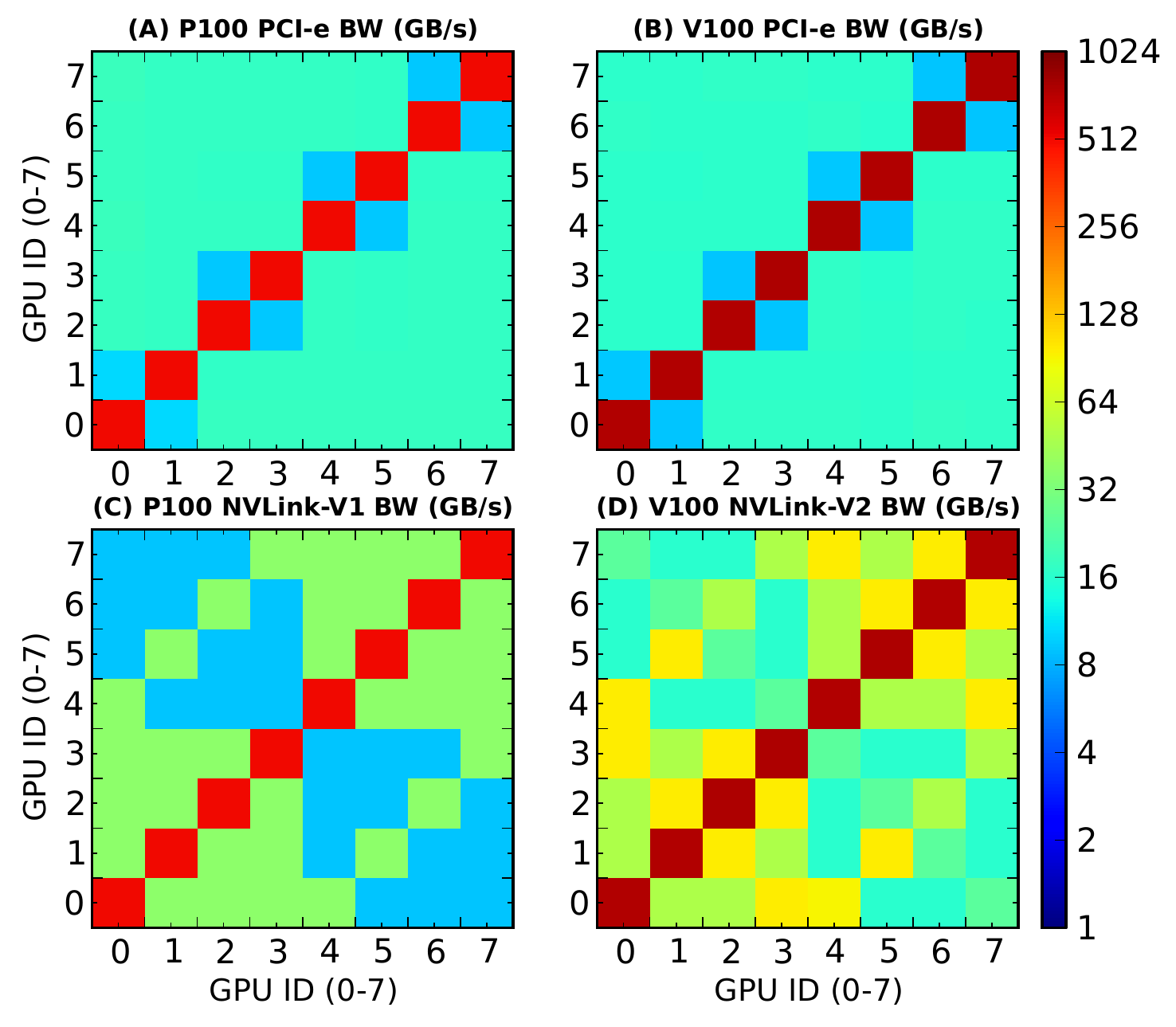} 
\caption{DGX-1 P2P bidirectional bandwidth. The 2x2 blue/red blocks along anti-diagonal in (A) and (B) clearly illustrate the anti-locality NUMA effect on PCIe. In (D), the yellow blocks compared with the green blocks show the NUMA among neighboring nodes. Meanwhile, the light green blocks along the anti-diagonal (not quite obvious though) imply the existence of NUMA among remote nodes.}
\label{fig:bibandwidth}
\endminipage
\end{figure*}

\section{GPU Interconnect Microbenchmarking}
\label{sec_evaluation}


We evaluate the basic characteristics of the six GPU interconnects using the microbenchmarks (listed in Table~\ref{tab:micro}) from the \emph{Tartan Benchmark Suite} \cite{li2018tartan} on the platforms listed in Table~\ref{tab:platform}, focusing on both \textbf{Peer-to-Peer (P2P)} and \textbf{Collective (CL)} communication patterns. For intra-node P2P, we especially concentrate on assessing the new node-level NVLink, NVSwitch and NV-SLI technologies in terms of topology, latency, bandwidth, efficiency on message size and NUMA effects. For inter-node P2P, we discuss properties such as latency, bandwidth and efficiency on message size. We use \emph{cudaEvent} to measure the latency and calculate the bandwidth.

\subsection{Intra-Node P2P Communication}

\subsubsection{Start-up Latency}
Figure~\ref{fig:latency} shows the start-up communication latency (i.e., raw latency for transmitting the shortest message) among arbitrary pair of GPUs via PCIe and NVLink for the P100 and V100 DGX-1 platforms.  As already mentioned, NVLink is not P2P self-routed; for GPUs that are not directly connected by NVLink (e.g., G0 and G5 in Figure~\ref{fig:dgx-1}), there are two routing paths that only require a single transit (e.g., from G0 to G5, either G1 or G4 can be the router). In such scenarios, Figure~\ref{fig:latency} shows the path exhibiting shorter latency. 

\vspace{4pt}\noindent \textbf{PCIe Latency:} Figure~\ref{fig:latency}-(A) and (B) demonstrate that the communication latency for accessing different pairs of GPUs via PCIe are similar, implying that no NUMA effects appear in latency through PCIe. In other words, the three types of latency by going through one PCIe switch (e.g., G0 and G1), across local CPU socket (e.g., G0 and G2), and across the QPI bridge (e.g., G0 and G6)  in Figure \ref{fig:dgx-1} are roughly the same. Meanwhile, comparing Figure~\ref{fig:latency}-(A) and (B), the PCIe latency is increased slightly (e.g., from green to light blue) from P100-DGX-1 to V100-DGX-1. As the bandwidth keeps unchanged, this may suggest a deeper communication pipeline design in V100-DGX-1 with Little's Law \cite{little1961proof}.

\vspace{4pt}\noindent \textbf{NVLink V1\&V2 Latency:} Compared to PCIe in Figure~\ref{fig:latency}-(A) and (B), NVLink in Figure~\ref{fig:latency}-(C) and (D) shows significant NUMA effects. For nodes that are directly connected, the latency is around 9$\mu s$; for nodes that require manual routing, the latency is increased by about 2x for P100-DGX-1 and 3x for V100-DGX-1. Again, we observe that NVLink-V2 exhibits higher latency than NVLink-V1, potentially due to a deeper pipeline or lower operating frequency.

\begin{figure*}[!htb]
\minipage{0.65\columnwidth}
\includegraphics[width=1.05\columnwidth]{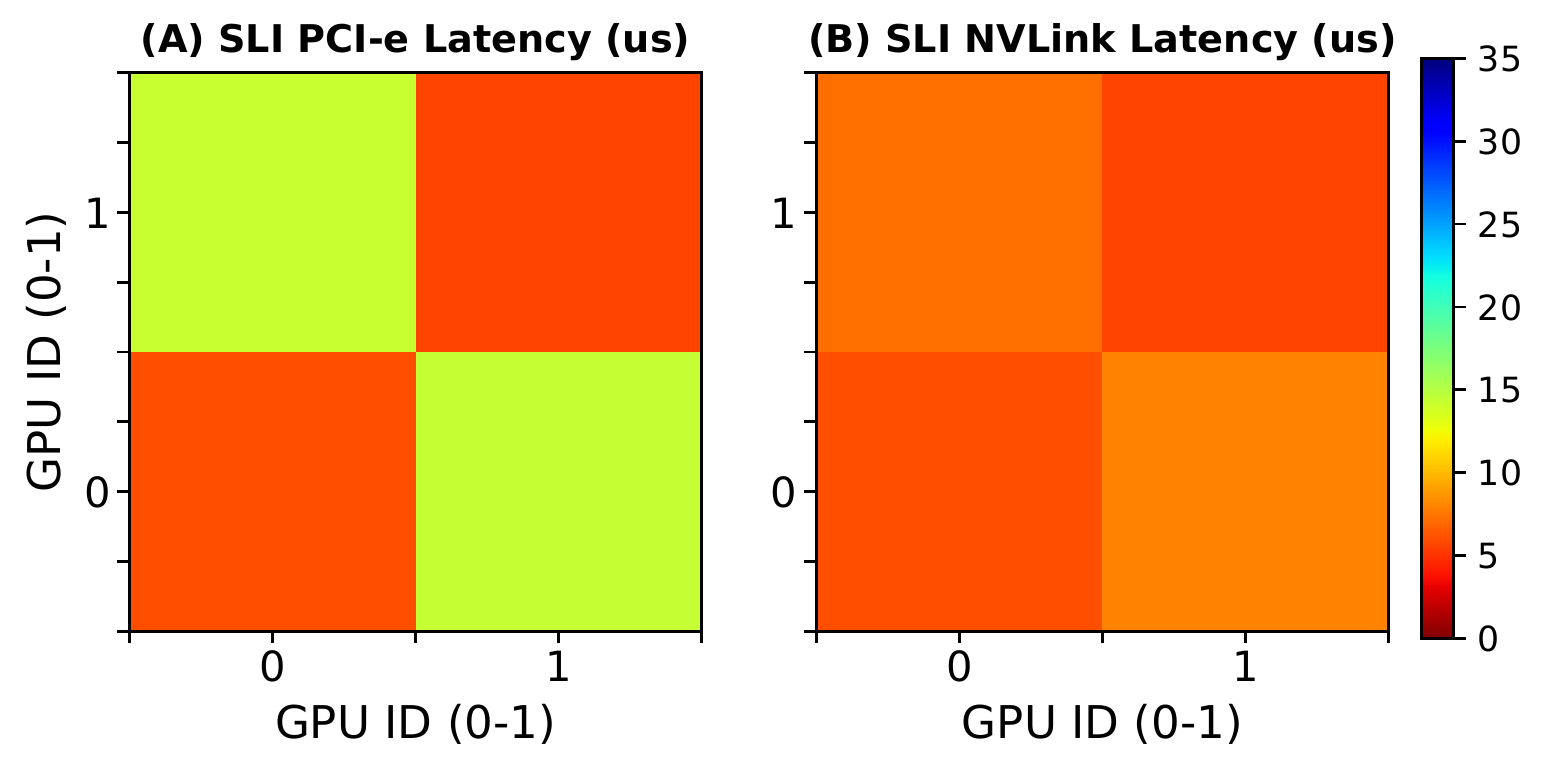} 
\caption{SLI-System P2P communication latency.}
\label{fig:latency-sli}
\endminipage\hfill
\minipage{0.66\columnwidth}
\includegraphics[width=1.05\columnwidth]{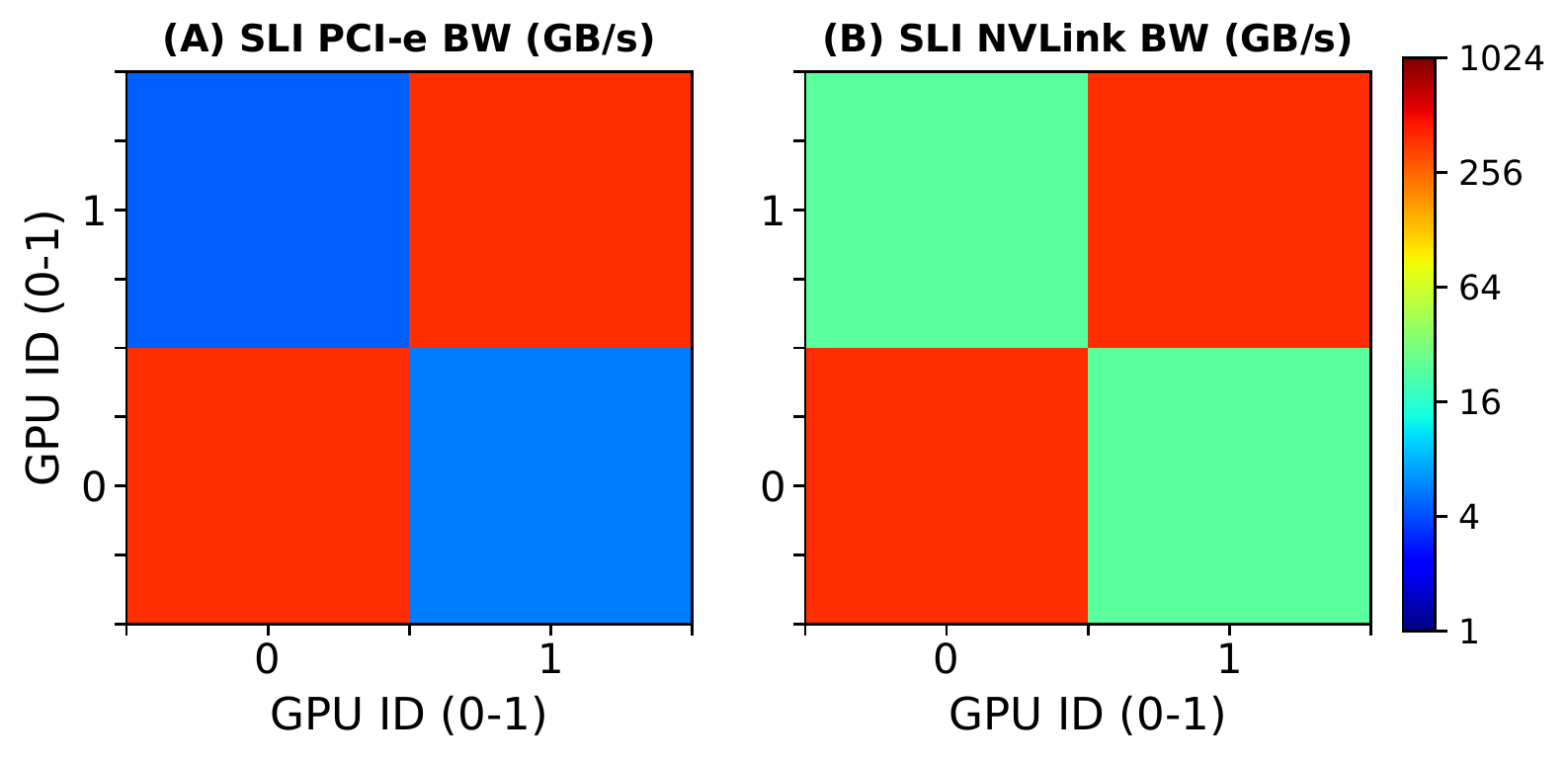} 
\caption{SLI-System P2P unidirectional bandwidth.}
\label{fig:unibandwidth-sli}
\endminipage\hfill
\minipage{0.66\columnwidth}
\includegraphics[width=1.05\columnwidth]{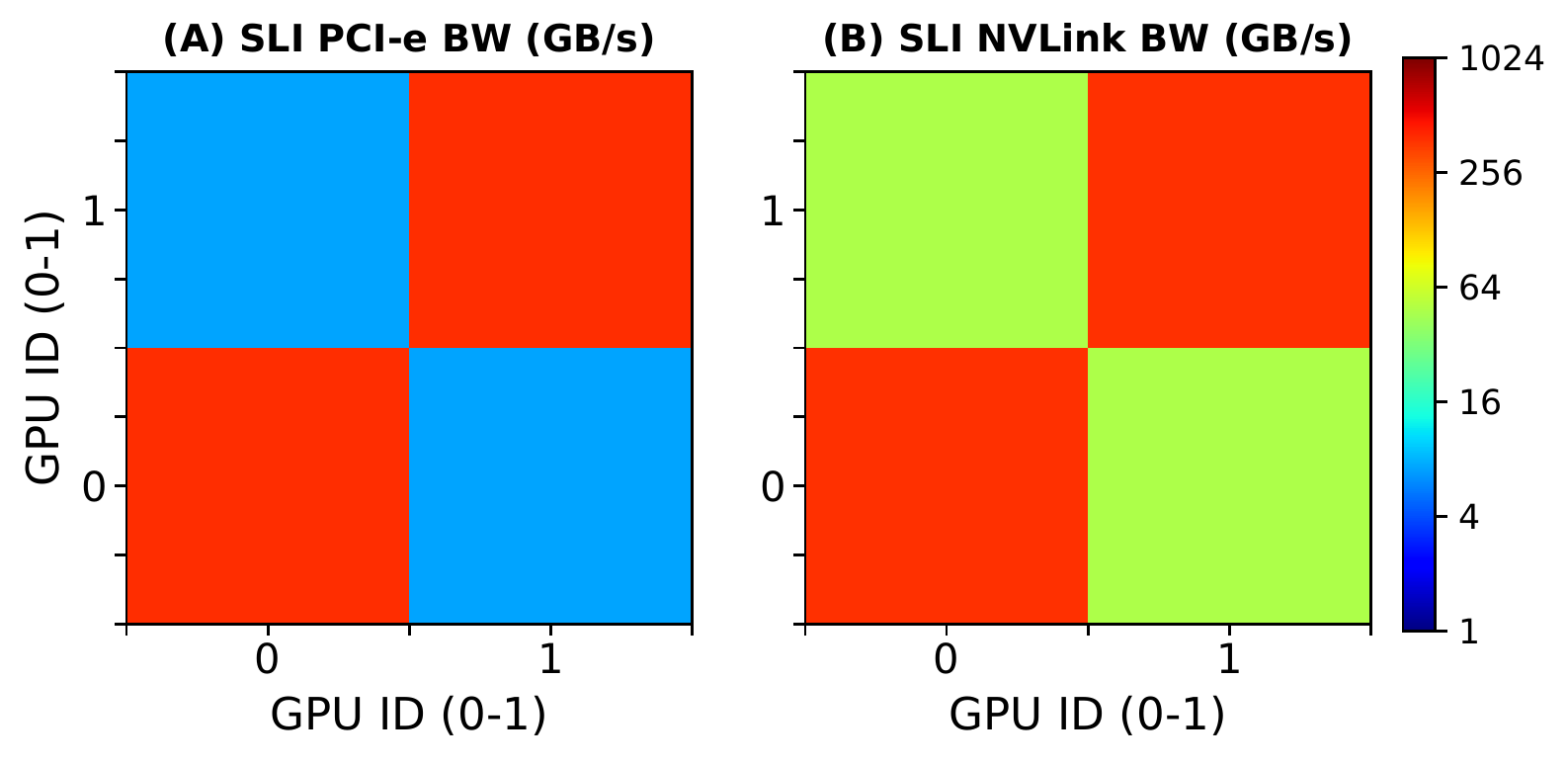} 
\caption{SLI-System P2P bidirectional bandwidth.}
\label{fig:bibandwidth-sli}
\endminipage
\end{figure*}

\begin{figure*}[!htb]
\minipage{0.65\columnwidth}
\includegraphics[width=1.05\columnwidth]{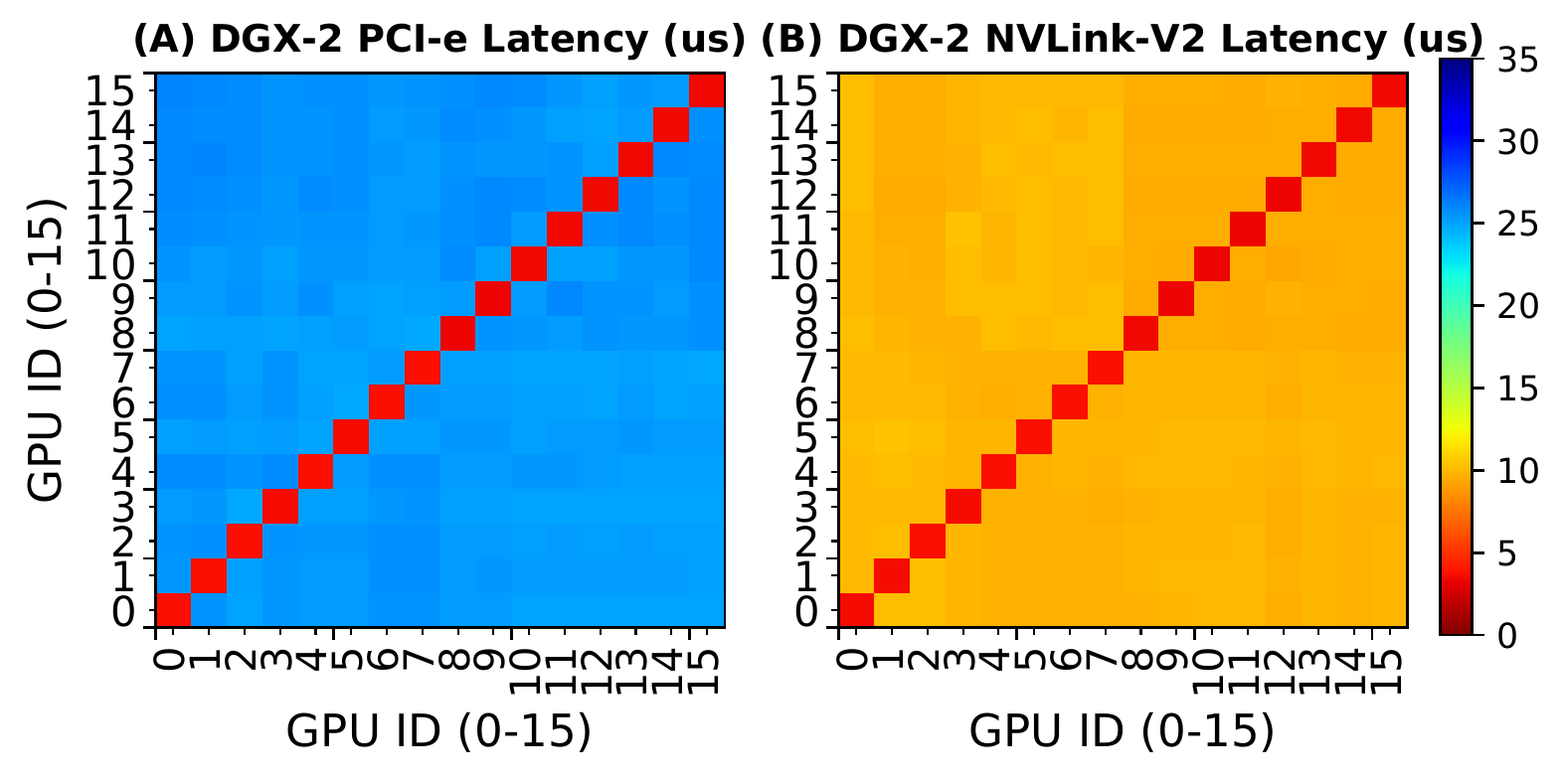} 
\caption{DGX-2 P2P communication latency.}
\label{fig:latency-nvswitch}
\endminipage\hfill
\minipage{0.66\columnwidth}
\includegraphics[width=1.05\columnwidth]{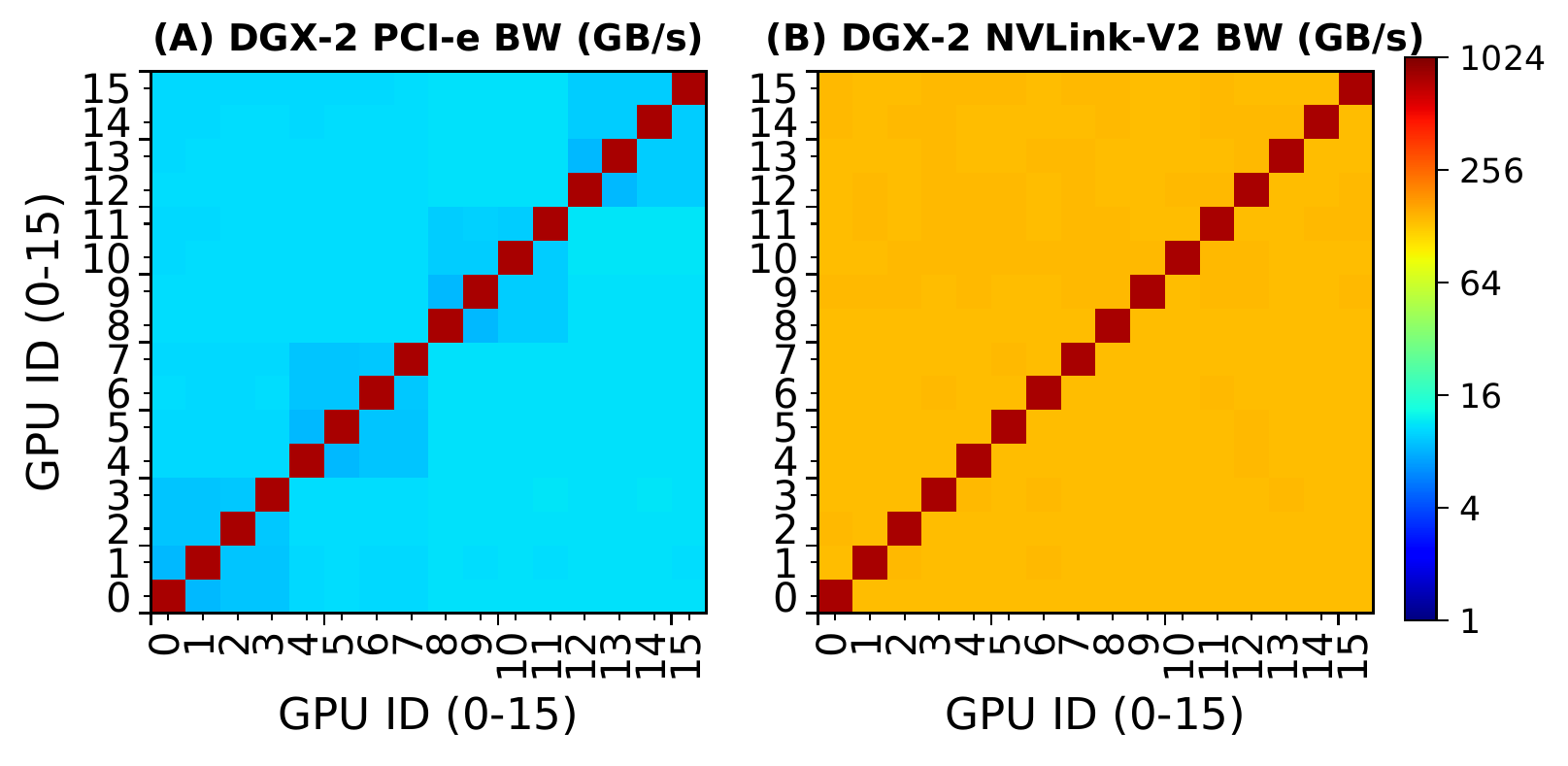} 
\caption{DGX-2 P2P unidirectional bandwidth.}
\label{fig:unibandwidth-nvswitch}
\endminipage\hfill
\minipage{0.66\columnwidth}
\includegraphics[width=1.05\columnwidth]{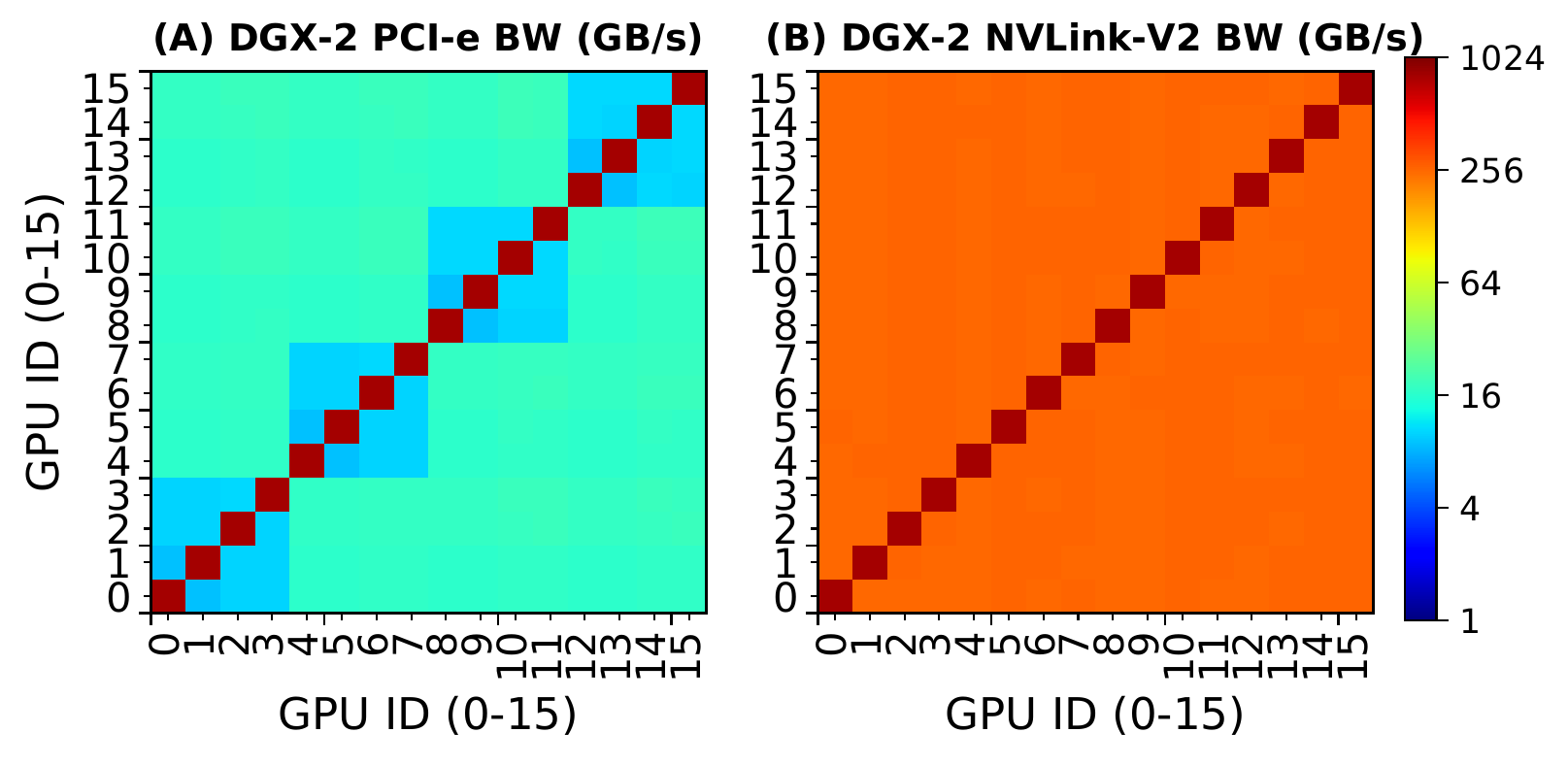} 
\caption{DGX-2 P2P bidirectional bandwidth.}
\label{fig:bibandwidth-nvswitch}
\endminipage
\end{figure*}

\vspace{4pt} \noindent \textbf{NV-SLI Latency:} Figure~\ref{fig:latency-sli} shows the latency for PCIe and NV-SLI in the SLI platform. For the dual-GPU system, there are three latency levels: \emph{local access} which is about 5$\mu$s; \emph{NV-SLI access} to the opposite GPU which is about 8$\mu$s; and \emph{PCIe access} to the opposite GPU, which is about 13$\mu$s. Note, the two GPUs are directly linked by NV-SLI without any intervention from other units such as CPU, DMA, etc.

\vspace{4pt} \noindent \textbf{NVSwitch Latency:} Figure~\ref{fig:latency-nvswitch} shows the latency for PCIe and NVSwitch in the DGX-2 platform. There are in total 16 GPUs in the system, so the grid is much finer. The pattern is very regular: all the remote access are homogeneous, either for PCIe or NVSwitch, confirming that NVSwitch is all-to-all fully-connected. Although accessing GPUs on the other baseboard incurs two switch hops, we can see that the difference of latency is very small, almost negligible.

\subsubsection{Sustainable Bandwidth}
Figure~\ref{fig:unibandwidth} and \ref{fig:bibandwidth} illustrate the uni- and bidirection sustainable bandwidth for PCIe and NVLink of the two DGX-1 platforms, respectively.  Figure~\ref{fig:unibandwidth-sli} and \ref{fig:bibandwidth-sli} show the uni- and bidirection bandwidth for PCIe and NV-SLI in the SLI platform. Finally, Figure~\ref{fig:unibandwidth-nvswitch} and Figure~\ref{fig:bibandwidth-nvswitch} show the uni- and bidirection bandwidth for PCIe and NVSwitch in DGX-2.

\vspace{4pt}\noindent \textbf{PCIe Unidirection Bandwidth:} From Figure~\ref{fig:unibandwidth}-(A) and (B), we can observe slight NUMA effects on PCIe accesses: two GPUs sharing the same PCIe switch (e.g., G2 and G3 in Figure~\ref{fig:dgx-1}) exhibit lower bandwidth in the measurement. For other GPUs, no matter whether sharing the same socket, the bandwidth appears to be the same. Similar effects have also been observed in Figure~\ref{fig:unibandwidth-nvswitch}-(A), in which four GPUs sharing the same Level-2 PCIe switch (e.g., G0 to G3 in Figure~\ref{fig:nvswitch-pcie}) deliver lower bandwidth.

\vspace{4pt}\noindent \textbf{PCIe Bidirection Bandwidth:} The NUMA effects on bidirectional bandwidth for GPUs sharing the same PCIe switch  (e.g., Figure~\ref{fig:bibandwidth}-(A) and (B), Figure~\ref{fig:bibandwidth-nvswitch}-(A)) are much more prominent than those on unidirection bandwidth. The PCIe NUMA effect here is an interesting novel observation: \emph{it describes a scenario that nearby access presenting lower performance than remote access.} We label such a NUMA effect as ``\textbf{anti-locality}''. To the best of our knowledge, few existing work have discussed this phenomenon before, without mentioning leveraging it for performance tuning practice. The anti-locality effect is possibly due to the unbalanced physical signal paths on the PCIe-switch chipsets \cite{antilocality}. Note, this PCIe anti-locality effect is only observed for bandwidth.

\vspace{4pt}\noindent \textbf{NVLink Unidirection Bandwidth:} The NVLink scenario is more complicated. For NVLink-V1 in Figure~\ref{fig:unibandwidth}-(C), there are three connection types: \emph{local access}, \emph{neighboring nodes} directly connected by NVLink, and \emph{remote nodes} requiring additional routing,  corresponding to the topology illustrated in Figure~\ref{fig:dgx-1}-(A). For NVLink-V2 in Figure~\ref{fig:unibandwidth}-(D), there are four connection types: local access, close neighboring nodes connected by dual links (i.e., the ``backbone ring"), general neighboring nodes connected by one link, and remote nodes, corresponding to the topology in Figure~\ref{fig:dgx-1}-(B). As such, there are three types of NUMA effects for NVLink:
\begin{itemize}[leftmargin=*]
 \item NUMA among neighbors and remote nodes for NVLink-V1 and V2 on both latency and bandwidth.
 \item NUMA among neighbor nodes for NVLink-V2. This is due to different number of links (either 1 or 2) to connect neighboring nodes in V100-DGX-1. Typically, the latency remains similar but these two types of links introduce bandwidth difference. 
 \item NUMA among remote nodes for NVLink-V2. This is caused by the choice of routing. Figure~\ref{fig:unibandwidth}-(C) and (D) show the bandwidth disparity for choosing different paths. 
\end{itemize}

\vspace{4pt}\noindent \textbf{NVLink Bidirection Bandwidth:} The three types of NVLink NUMA effects for bidirectional bandwidth are much more obvious than that for unidirectional bandwidth, as discussed in the caption of Figure~\ref{fig:bibandwidth}.

\vspace{4pt}\noindent \textbf{NV-SLI Unidirection Bandwidth:} Since NV-SLI only incorporates two GPUs, where the communication is symmetric, showing no NUMA effect in Figure~\ref{fig:unibandwidth-sli}-(B). 

\vspace{4pt}\noindent \textbf{NV-SLI Bidirection Bandwidth:} The bidirectional condition is similar to unidirection condition, except that the bandwidth doubles,  as shown in Figure~\ref{fig:bibandwidth-sli}-(B). 


\vspace{4pt}\noindent \textbf{NVSwitch Unidirection Bandwidth:} Shown in Figure~\ref{fig:unibandwidth-nvswitch}-(B), the bandwidth for all remote access through NVSwitch are consistent or UMA, implying that one more NVSwitch hop does not degrade bandwidth. 

\vspace{4pt}\noindent \textbf{NVSwitch Bidirection Bandwidth:} Bidirection bandwidth condition is similar, except that the bandwidth doubles, as shown in Figure~\ref{fig:bibandwidth-nvswitch}-(B).

\begin{figure}[!t]
\centering
\includegraphics[width=\columnwidth]{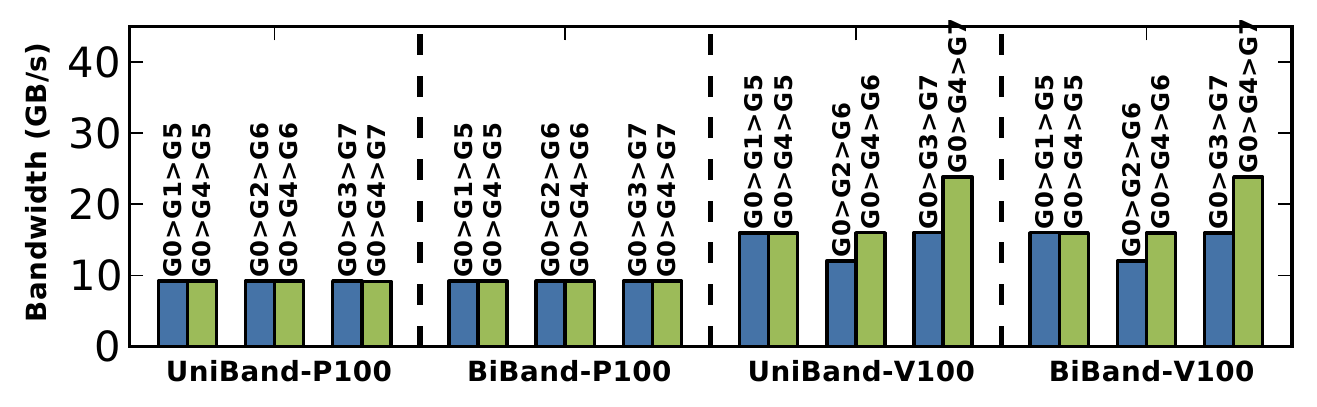} 
\caption{NUMA effect on routing choices for remote GPU access via NVLink.}
\label{fig:numa}
\end{figure}

\subsubsection{Routing} 

For all the GPU interconnects we discuss here, only the one for remote access in the DGX-1s via NVLink may require explicit routing. Here, we further explore the NUMA effects on alternative routing choices. For demonstration purposes, we take G0 in Figure~\ref{fig:dgx-1} as the source node for P2P communication. There are three remote nodes for G0: G5, G6 and G7. From G0 to G5, either G1 or G4 can be specified for routing. From G0 to G6, either G2 or G4 can be selected; and from G0 to G7, either G3 or G4 can be selected. We use microbenchmarks from Tartan to measure the latency, unidirection bandwidth and bidirection bandwidth of each routing path from G0 to G5, G6, G7 respectively on both DGX-1 platforms. Figure~\ref{fig:numa} shows the results for unidirection and bidirection bandwidth. As can be seen, for NVLink-V1 in P100-DGX-1, there are no NUMA effects; all the bars appear in the same height. This is because the NVLinks in P100-DGX-1 are isomorphic --- any connection incorporates just one link. However, for NVLink-V2 in V100-DGX-1, different scenarios emerge based on how many dual-bandwidth links a route goes through, e.g., the lowest bandwidth occurs from G0$\rightarrow$G2$\rightarrow$G6 while the highest is triggered by routing G0$\rightarrow$G4$\rightarrow$G7. Nevertheless, the latency remains similar for all scenarios (not shown in the figures).



\subsubsection{Efficiency on Message Size}

\begin{figure*}[!t]
\centering
\includegraphics[width=2\columnwidth]{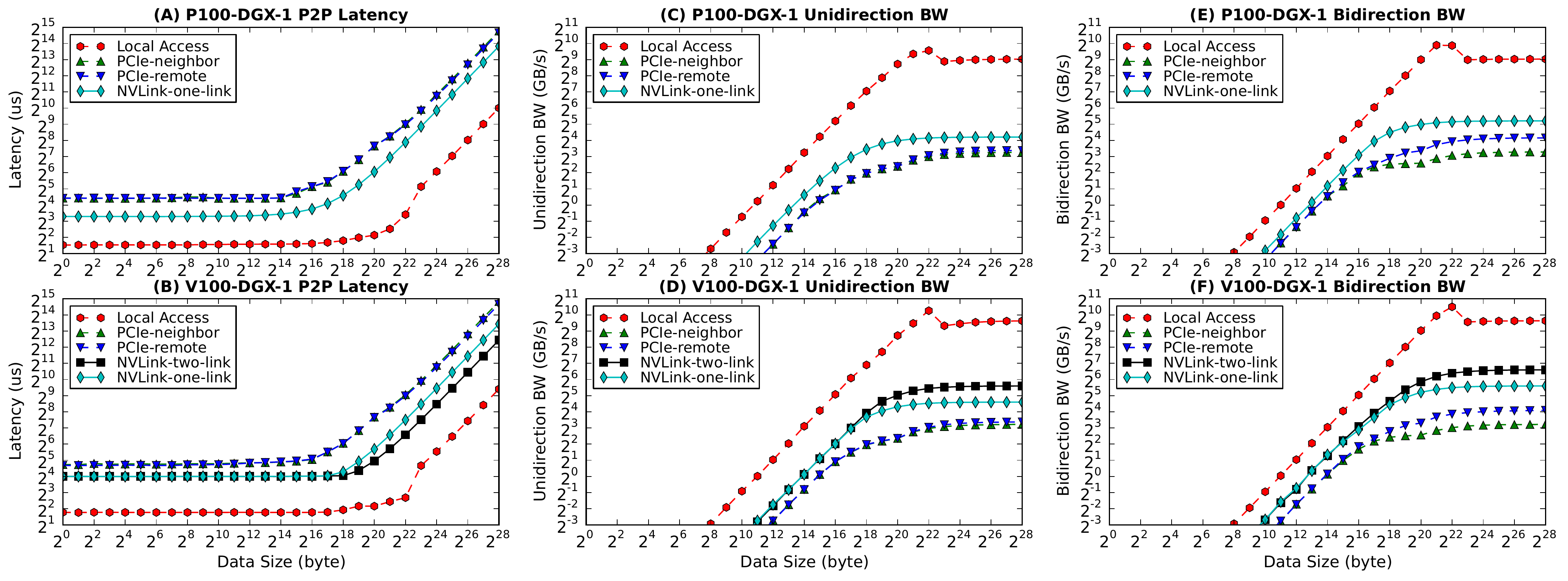} 
\caption{P2P communication latency, unidirection and bidirection bandwidth with increased message size via PCIe and NVLink for DGX-1.} 
\label{fig:packet}
\end{figure*}

\begin{figure*}[!t]
\centering
\includegraphics[width=2\columnwidth]{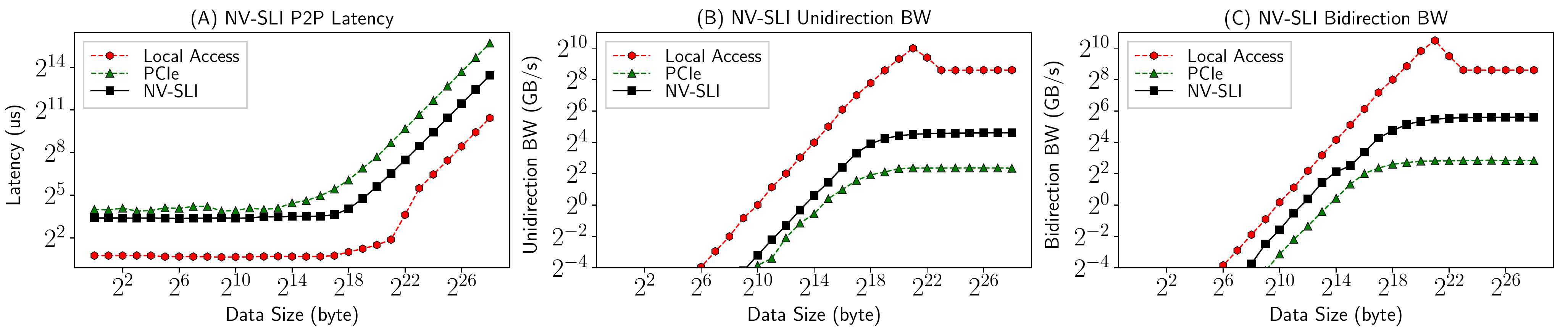} 
\caption{P2P communication latency, unidirection and bidirection bandwidth with increased message size via PCIe and NV-SLI for the SLI-system.} 
\label{fig:packet-sli}
\end{figure*}

\begin{figure*}[!t]
\centering
\includegraphics[width=2\columnwidth]{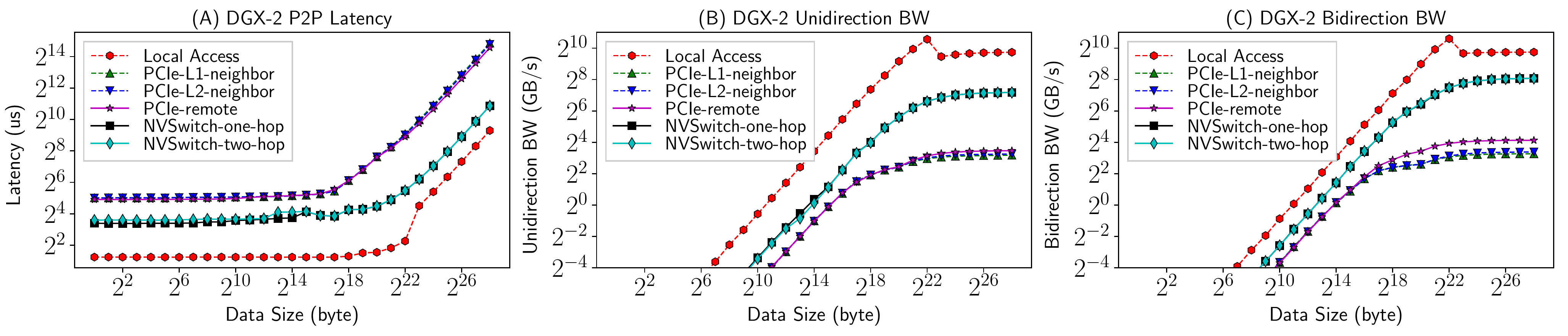} 
\caption{P2P communication latency, unidirection and bidirection bandwidth with increased message size via PCIe and NVSwitch for DGX-2.} 
\label{fig:packet-nvswitch}
\end{figure*}

Figure~\ref{fig:packet} illustrates the P2P latency, unidirection and bidirection bandwidth with respect to message size via PCIe, NVLink-V1 and V2 for P100-DGX-1 and V100-DGX-1. Figure~\ref{fig:packet-sli} illustrates the plot for NV-SLI and PCIe in the SLI-system. Figure~\ref{fig:packet-nvswitch} illustrates the plot for NVSwitch and PCIe in the DGX-2 system. 


\vspace{4pt}\noindent \textbf{Latency:} The latency remains unchanged for data communication less than or equal to 16KB for P100-DGX-1 (except local access). For V100-DGX-1, this value increases to 64KB, suggesting higher link bandwidth to saturate and deeper communication pipeline on NVLink-V2. The conditions are similar for SLI and NVSwitch in the SLI and DGX-2 platforms. The interesting observation is that in Figure~\ref{fig:packet-nvswitch}, there is slight divergence for PCIe-local (including PCIe-neighbor and PCIe-one-switch) and PCI-remote access latency with large messages (i.e., $\ge$4MB). 



\vspace{4pt}\noindent \textbf{Bandwidth:} For unidirection bandwidth, it can be seen that the interconnect starts to saturate at about $2^{22}=$4MB for both NVLink-V1 and V2, implying that to reach the optimal sustainable bandwidth of the interconnect, one needs at least 4MB data to transmit at a time. This is also true for bidirection bandwidth in Figure~\ref{fig:packet}-(E) and (F). In addition, observing the fact that the latency starts to increase at 16KB and 64KB, but the bandwidth begins to saturate at 4MB, implies that from 64KB, we suffer from extra delay, but still gain overall bandwidth improvement until the bandwidth saturation point. For DGX-2 NVSwitch uni- and bidirection bandwidth in Figure~\ref{fig:packet-nvswitch}-(B) and (C), we see that NVSwitch-one-hop and NVSwitch-two-hop curves are fully aligned, confirming that accessing remote baseboard imposes no extra overhead. For PCIe bidirection bandwidth in Figure~\ref{fig:packet-nvswitch}-(C), the observation is that when message size is larger than 64KB, the PCIe anti-locality effect appears: PCIe-remote access delivers higher bidirection bandwidth than PCIe-neighbors (L1 \& L2). In fact, the anti-locality effect also exists between L1-PCIe-neighbors and L2-PCIe-neighbor in DGX-2, as can be seen, the bandwidth of L2-PCIe-neighbor is slightly better than L1-PCIe-neighbor, showing a second level anti-locality. Finally, all GPU local access bandwidth in these figures staggers at about 4MB, possibly due to exceeding page boundary.





\subsection{Intra-Node Collective Communication}

\label{subsec_intracl}
Different from P2P communication only including a single sender and receiver, collective communication (CL) involves multiple senders and receivers so it is more complicated. CL generally follows certain patterns, including \emph{broadcast, scatter, gather, all-gather, reduce, all-reduce, all-to-all}, etc. It is extensively used in many key applications such as deep learning, molecular dynamics, graph analytics, etc.

Efficiently implementing CL communication is challenging because (a) it needs to understand the underlying hardware network topology in order to enable orchestrated mapping; (b) it needs to handle the issue of synchronization, overlapping and deadlock; and (c) performance metrics can differ subject to application features (e.g., latency-oriented for small transfers but bandwidth-oriented for large transfers). To relieve these burdens from users, NVIDIA provides Collective Communication Library (NCCL) \cite{ncclV1, ncclV2}, using similar primitives as MPI collectives, while AMD offers RCCL \cite{rccl}. NCCL currently supports five CL patterns: \emph{broadcast, all-gather, reduce, all-reduce}, and \emph{reduce-scatter}.

\begin{figure}[!t]
\centering
\includegraphics[width=\columnwidth]{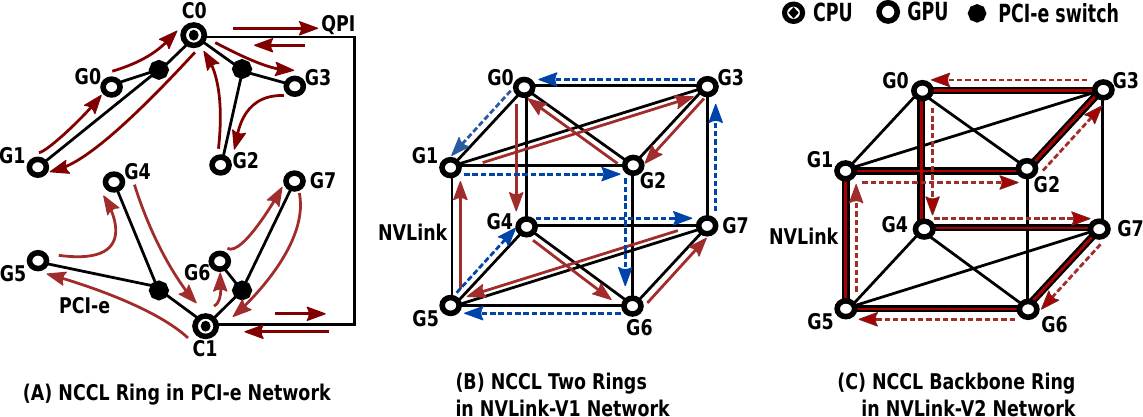} 
\caption{NCCL Rings for PCIe, NVLink-V1 and NVLink-V2 interconnect. (A) for PCIe, the ring is to traverse the binary-tree network; (B) for NVLink-V1, there are two independent rings, marked in red-solid line and blue-dash line; and (C) for NVLink-V2, the lines with 2 links form a fast ring (i.e., the backbone network) while the lines with 1 link form a slow ring.}
\label{fig:nccl}
\end{figure}

To offer the maximum bandwidth, NCCL constructs ring network among the communication participants so that broadcasting and reduction can be efficiently realized by partitioning data into small chunks, and transmitting them along the ring in a pipeline fashion. It is claimed that the ring algorithm can provide near optimal bandwidth for most of the standard CL operations and can be easily applied to various network topology \cite{nvidia2017dgx1}. Figure~\ref{fig:nccl}, as an example, describes how a ring-network is constructed for PCIe, NVLink-V1 and NVLink-V2 in DGX-1s, respectively.

There are two versions of the NCCL library: NCCL-V1 \cite{ncclV1} is opensource but only supports intra-node PCIe/QPI interconnect network. NCCL-V2 \cite{ncclV1} is closed-source but supports NVLink, PCIe, NVSwitch, NV-SLI, IB and IP networks, and can automatically integrate them to maximize overall bandwidth. Although the combination improves overall performance, it also introduces difficulty in independently measuring the CL communication efficiency for a particular interconnect. Consequently, NCCL-V1 is employed for PCIe CL evaluation while NCCL-V2 is adopted for NVLink/NVSwitch/NV-SLI CL evaluation.

\subsubsection{DGX-1 CL Communication}

\begin{figure*}[!t]
\centering
\includegraphics[width=2\columnwidth]{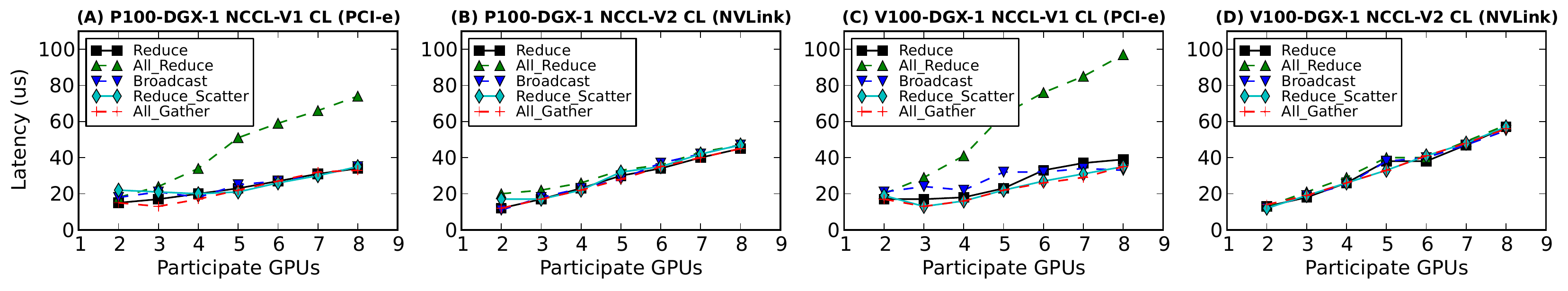} 
\caption{Intra-node CL communication latency with variable participant GPUs for NCCL-V1 (PCIe/QPI) and NCCL-V2 (NVLink-V1/2).}
\label{fig:nccl_latency}
\end{figure*}

\vspace{3pt}\noindent \textbf{CL Latency:} Figure~\ref{fig:nccl_latency} illustrates CL communication startup latency with respect to the number of GPUs involved for NCCL-V1 and V2 on the two DGX-1 platforms respectively, corresponding to PCIe/QPI and NVLink. We have the following observations: (1) latency increases with participating GPUs almost in a linear fashion; (2) comparing (A) and (C), (B) and (D), the behaviors of NCCL-V1 and V2 on the two DGX platforms are similar; (3) comparing (A) (B) with (C) (D), the latency of NVLink increases faster than PCIe (except all-reduce), while NVLink-V2 increases faster than NVLink-V1; (4) for PCIe, \emph{all\_reduce} shows the largest latency. The disalignment of the curves in (B) and (D) for odd number of GPUs (e.g., 3, 5) is likely due to NCCL algorithm design rather than NVLink-V2 P2P NUMA effects, as will be discussed later.

\begin{figure*}[!t]
\centering
\includegraphics[width=2\columnwidth]{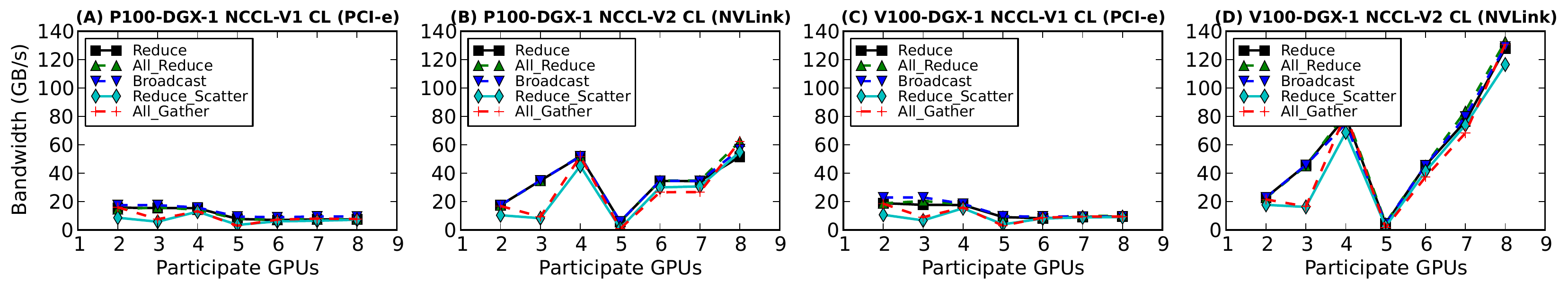} 
\caption{Intra-node CL communication bandwidth with variable participant GPUs for NCCL-V1 (PCIe/QPI) and NCCL-V2 (NVLink-V1/2).}
\label{fig:nccl_bandwidth}
\end{figure*}

\begin{figure*}[!t]
\centering
\includegraphics[width=2\columnwidth]{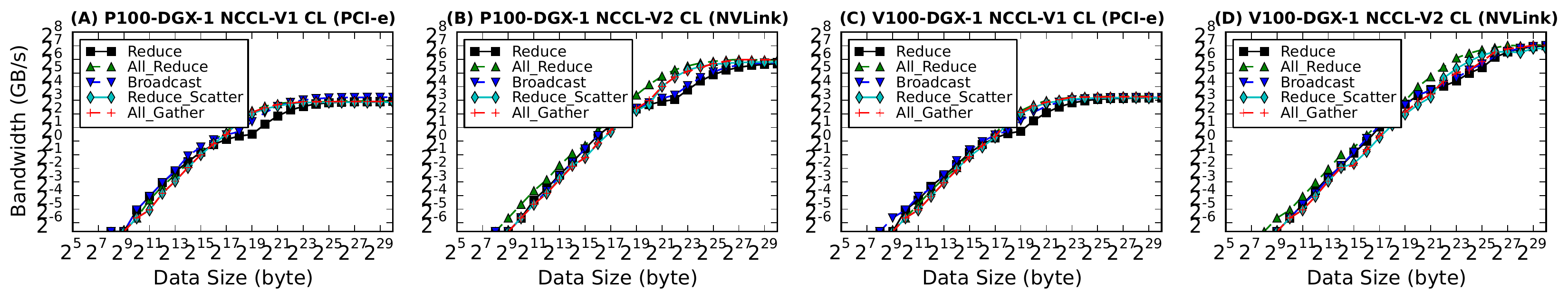} 
\caption{Intra-node CL communication bandwidth for 8 GPUs with increased message size for NCCL-V1 (PCIe/QPI) and NCCL-V2 (NVLink-V1/2).}
\label{fig:nccl_packet}
\end{figure*}

\vspace{3pt}\noindent \textbf{CL Bandwidth:} Figure~\ref{fig:nccl_bandwidth} shows CL's sustainable communication bandwidth with increased number of GPUs under 1GB payload. As can be seen, (1) for PCIe, CL bandwidth decreases with more GPUs, which is due to bus contention in a tree-network, see Figure~\ref{fig:nccl}-(A); (2) for NVLink, however, CL bandwidth essentially increases with more GPUs due to more connected links in a hypercube mesh network, see Figure~\ref{fig:nccl}-(B) and (C); (3) PCIe and NVLink behavior on P100-DGX-1 and V100-DGX-1 are in similar trend. However, NVLink-V2 exhibits significantly better bandwidth with 4 GPUs ($\sim$1.6x) and 8 GPUs ($\sim$2x) compared to NVLink-V1, showing the strength of dual-links and backbone ring (Figure~\ref{fig:nccl}); (4) the NUMA effects appear quite significant with 5 GPUs, implying that there may be a congestion when forming a NCCL ring among 5 GPUs. One should avoid adopting 5 GPUs in their application setup.

\vspace{3pt}\noindent \textbf{CL Efficiency on Message Size:}  Figure~\ref{fig:nccl_packet} shows CL bandwidth with respect to message size increasing from 8B to 1GB for 8 GPUs. As can be seen, for PCIe, CL-bandwidth saturates at about $2^{24}=16$MB; whereas for NVLink, bandwidth saturates around $2^{28}=256$MB. Again, the five CL patterns exhibit similar trend in terms of bandwidth when scaling the message size.

\subsubsection{NV-SLI CL Communication}

\begin{figure*}[!htb]
\minipage{0.99\columnwidth}
\includegraphics[width=1.02\columnwidth]{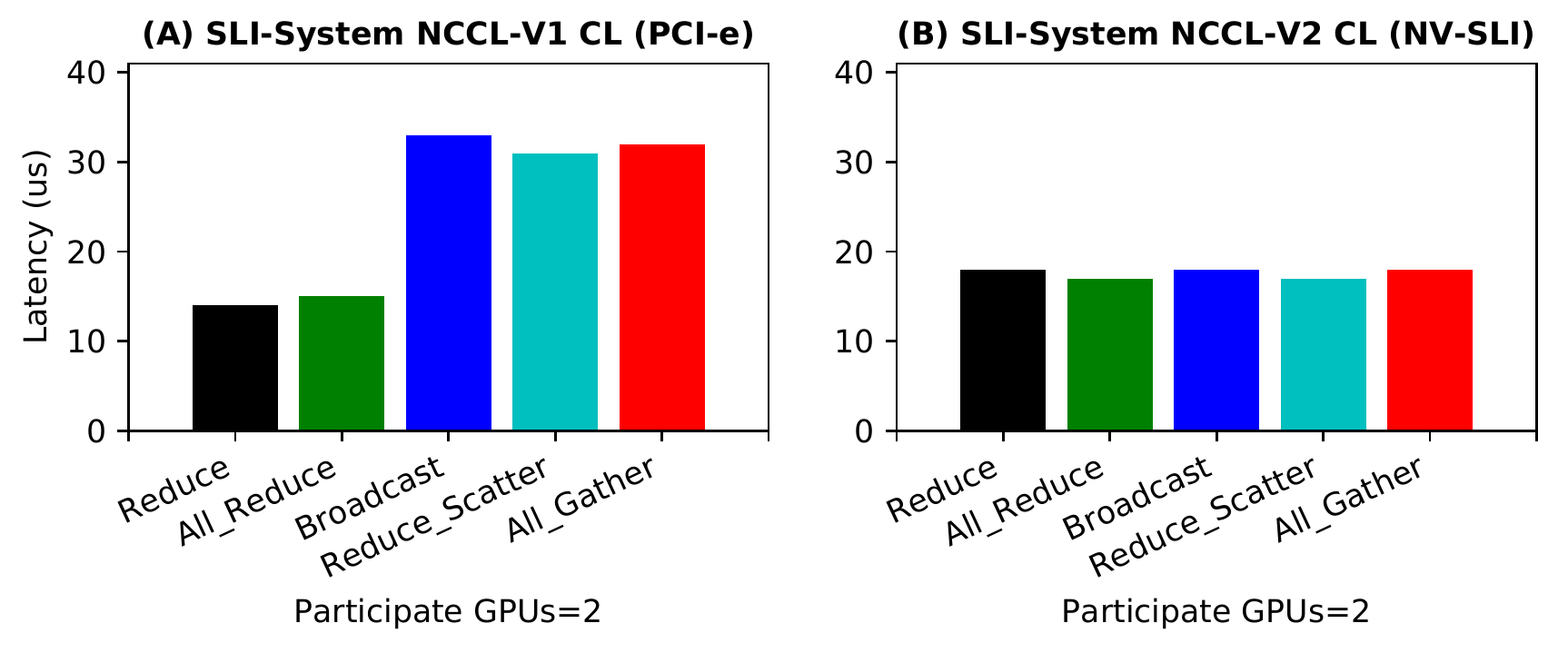} 
\caption{SLI-system PCIe and NV-SLI CL communication latency.}
\label{fig:sli_nccl_latency}
\endminipage\hfill
\minipage{0.99\columnwidth}
\includegraphics[width=1.02\columnwidth]{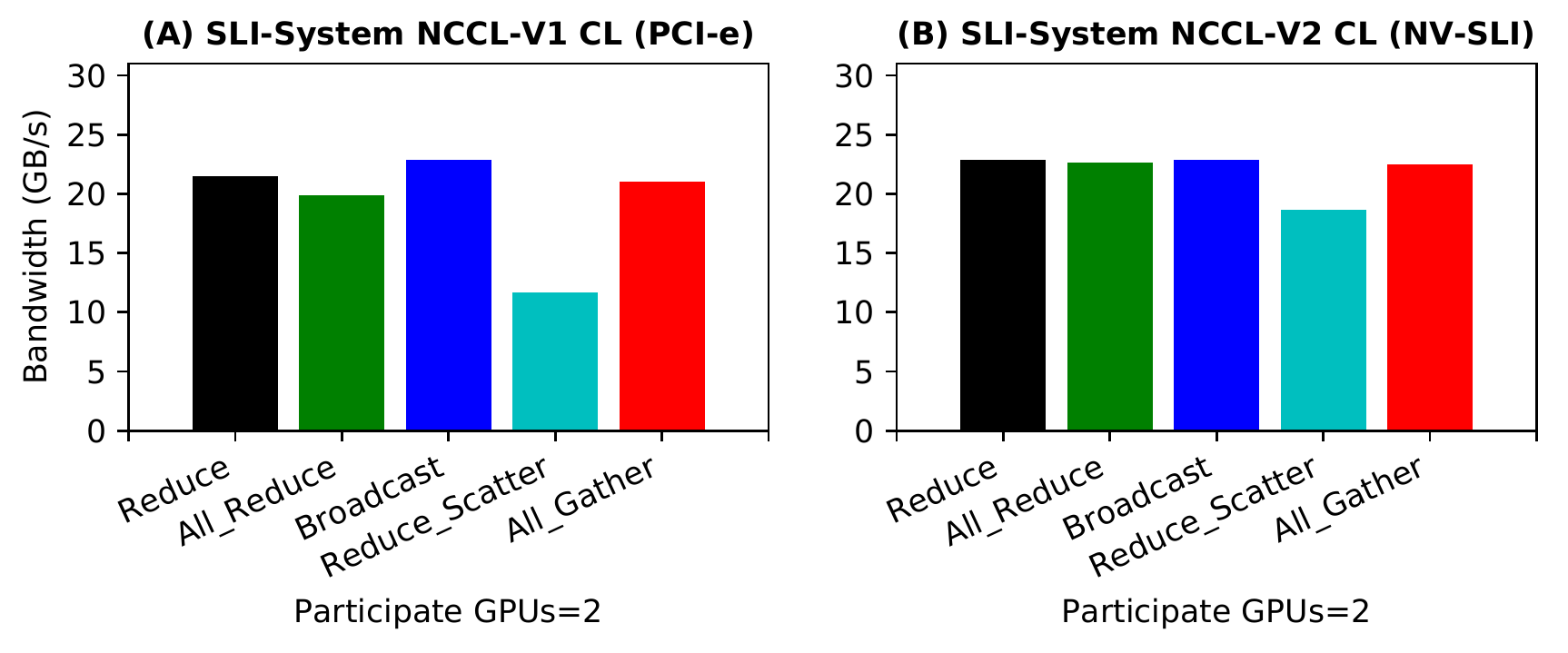} 
\caption{SLI-system PCIe and NV-SLI CL communication bandwidth.}
\label{fig:sli_nccl_bandwidth}
\endminipage
\end{figure*}

\vspace{3pt}\noindent \textbf{CL Latency and Bandwidth:} Since the SLI-system contains only two GPUs, we use histogram figures to show the latency and bandwidth for the 5 CL communication patterns. As can be seen in Figure~\ref{fig:sli_nccl_latency} and \ref{fig:sli_nccl_bandwidth}, the latency for NV-SLI is similar, around 18$\mu s$; but for PCIe, \emph{reduce} and \emph{all\_reduce} show significantly lower latency than the other three, even less than on NV-SLI. The bandwidth are similar for both PCIe and NV-SLI, except that \emph{reduce\_scatter} showing poorer bandwidth than the others on both PCIe and NV-SLI.

\begin{figure*}[!htb]
\minipage{0.99\columnwidth}
\includegraphics[width=1.02\columnwidth]{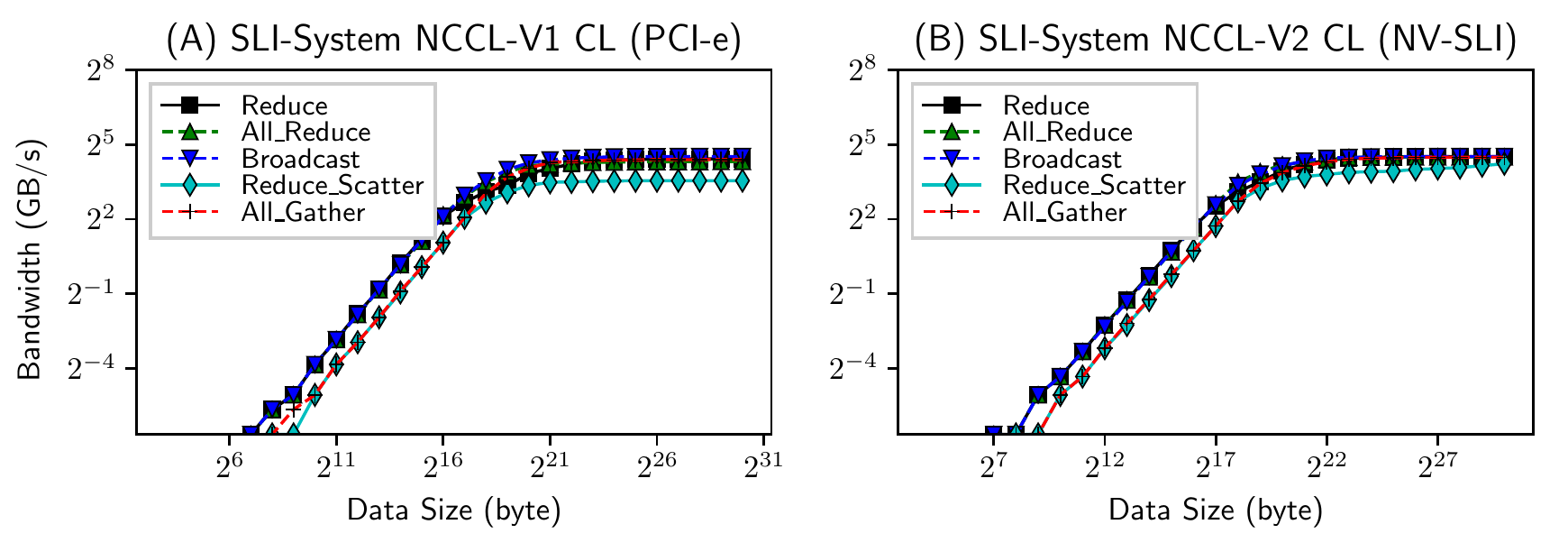} 
\caption{SLI-system PCIe and NV-SLI CL bandwidth efficiency.}
\label{fig:sli_nccl_packet}
\endminipage\hfill
\minipage{0.99\columnwidth}
\includegraphics[width=1.02\columnwidth]{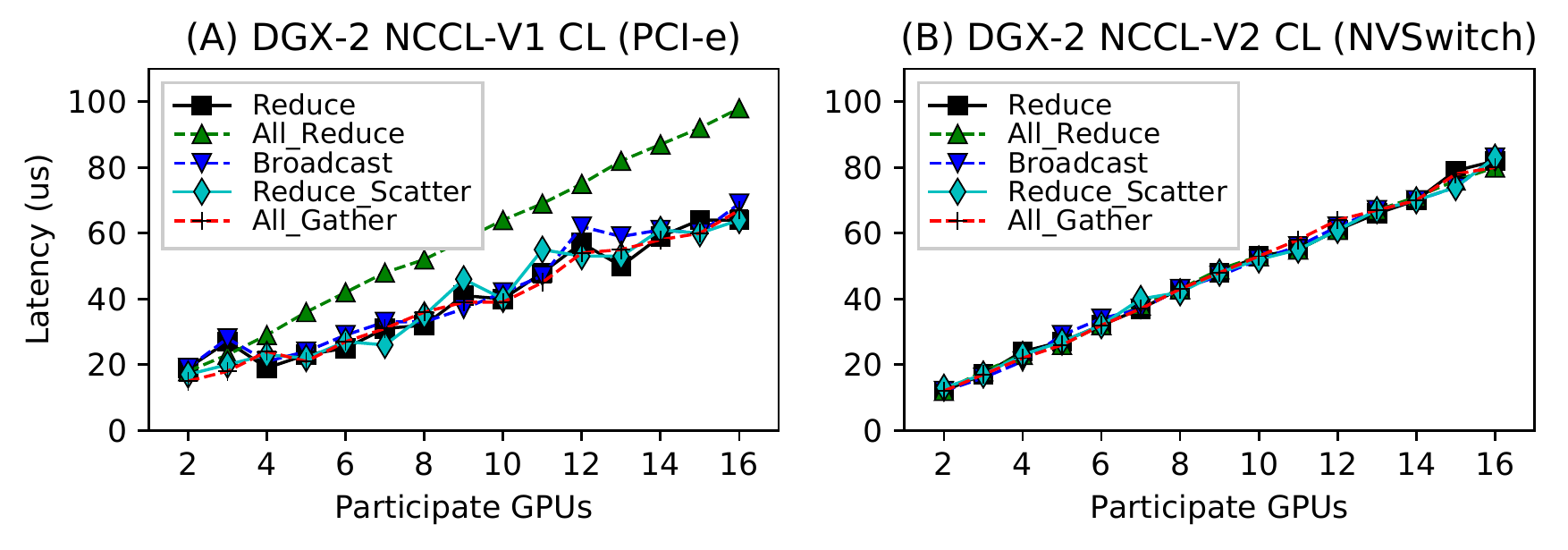} 
\caption{DGX-2 PCIe and NVSwitch CL communication latency.}
\label{fig:nvswitch_nccl_latency}
\endminipage
\end{figure*}

\vspace{3pt}\noindent \textbf{CL Efficiency on Message Size:} Figure~\ref{fig:sli_nccl_packet} shows CL bandwidth with respect to message size for the two GPUs. Both PCIe and NV-SLI bandwidth start to saturate at about $2^{20}=1$MB. The figure confirms that \emph{reduce\_scatter} bandwidth is lower than the others for large message size, but at the same time indicating that \emph{all\_gather} delivers a relative low bandwidth with the same message size before saturated.

\begin{figure*}[!htb]
\minipage{0.99\columnwidth}
\includegraphics[width=1.02\columnwidth]{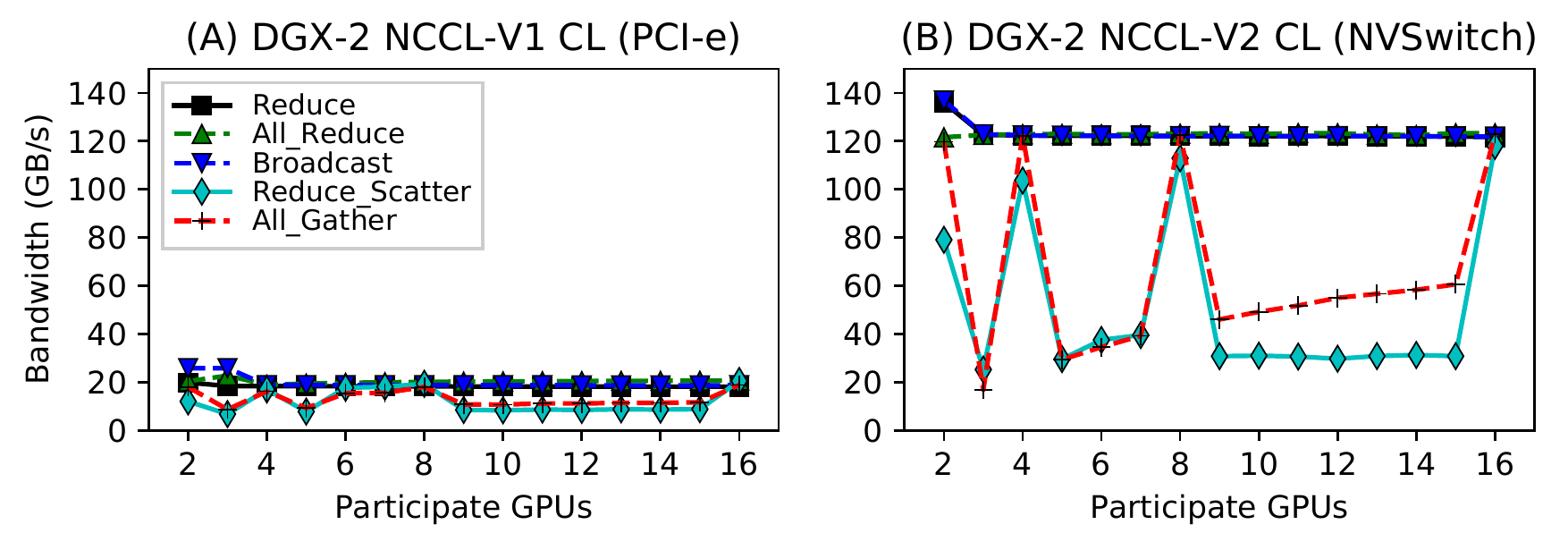} 
\caption{DGX-2 PCIe and NVSwitch CL communication bandwidth.}
\label{fig:nvswitch_nccl_bandwidth}
\endminipage\hfill
\minipage{0.99\columnwidth}
\includegraphics[width=1.02\columnwidth]{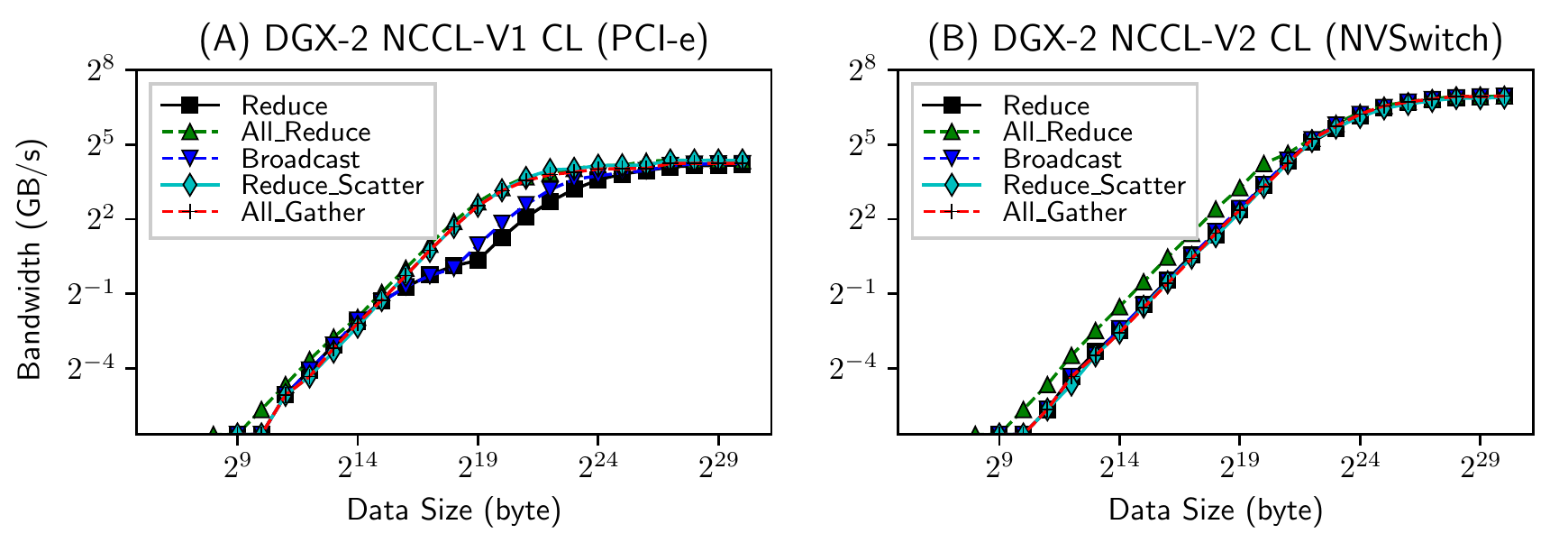} 
\caption{DGX-2 PCIe and NVSwitch CL bandwidth efficiency with 16 GPUs.}
\label{fig:nvswitch_nccl_packet}
\endminipage
\end{figure*}

\subsubsection{NVSwitch CL Communication}

\vspace{4pt}\noindent \textbf{CL Latency and Bandwidth:} Figure~\ref{fig:nvswitch_nccl_latency} illustrates CL communication latency with respect to the number of GPUs on DGX-2, corresponding to PCIe and NVSwitch. As can be seen, due to UMA for NVSwitch, the five curves in Figure~\ref{fig:nvswitch_nccl_latency}-(B) are rather aligned. Figure~\ref{fig:nvswitch_nccl_latency}-(B) confirms the long latency for \emph{all\_reduce} on tree-based PCIe interconnect, in consistent with Figure~\ref{fig:nccl_latency}-(A) and (C). Also note that except \emph{all\_reduce}, the other CL primitives show lower startup latency on PCIe than on NVSwitch when the participanting GPUs are more than three, implying that the advantage of NVSwitch (as well as NVLink) is on bandwidth rather than latency. This observation is confirmed in Figure~\ref{fig:nvswitch_nccl_bandwidth}. As can be seen, NVSwitch shows tremendously higher bandwidth than PCIe, particularly for \emph{reduce}, \emph{all\_reduce} and \emph{broadcast}. \emph{reduce\_scatter} and \emph{all\_gather} show staggering behavior on bandwidth for NVSwitch; the values are much better with 2, 4, 8 and 16 GPUs than other numbers of GPUs, in consistent with NVLink scenarios in Figure~\ref{fig:nccl_bandwidth}-(B)/(D). Since NVSwitch is UMA, it implies that this staggering issue is not due to interconnect topology but NCCL's implementation.  

\vspace{4pt}\noindent \textbf{CL Efficiency on Message Size:} Figure~\ref{fig:nvswitch_nccl_packet} shows CL bandwidth with respect to message size for 16 GPUs on the DGX-2 system. As can be seen, the curves for PCIe in Figure~\ref{fig:nvswitch_nccl_packet}-(A) are in consistent with PCIe in Figure~\ref{fig:nccl_packet}-(A) and (C): the five curves are mostly aligned for small (e.g., $\le$32KB) and large (e.g., $\ge$32MB) messages, but not the case in-between for \emph{reduce} and \emph{broadcast}. This divergence also appears for NVLink-V1 \& V2 in Figure~\ref{fig:nccl_packet}-(B) and (D), but disappears for NVSwitch in Figure~\ref{fig:nvswitch_nccl_packet}-(B). As NVSwitch is UMA, we suppose this disalignment is brought by the NUMA effect in the PCIe and NVLink networks.

\subsection{Inter-Node P2P Communication}
\label{subsec_interp2p}

We measure the latency and bandwidth of inter-node P2P communication on \emph{SummitDev Supercomputer} \cite{summitdev} and \emph{Summit Supercomputer} \cite{summit} from \emph{Oak Ridge National Laboratory}. SummitDev is a supercomputer system with 54 nodes. Each node contains two IBM Power-8 CPUs and four NVIDIA P100 GPUs. The GPU interconnect is NVLink-V1. Summit features 4,608 nodes, each with two IBM Power-9 CPUs and six NVIDIA V100 GPUs. The GPU interconnect is NVLink-V2. Both SummitDev and Summit support GPUDirect.

For inter-node P2P, we conduct our measurement under five configurations: (i) \emph{GPUDirect-RDMA} is to directly access GPU memory among nodes with GPUDirect enabled (``\emph{GPUDirect}'' here refers to a system option. On SummitDev, it is \emph{OMPI\_MCA\_pml\_pami\_enable\_cuda=1}. On Summit, it is \emph{--smpiargs=``gpu''.}) (ii) \emph{PinnedMem-GPUDirect} is to first copy data from GPU memory to the pinned host memory, then transfer the data to another node's pinned host memory and finally copy to the targeted GPU memory, with GPUDirect enabled; (iii) \emph{PinnedMem} is similar to (ii) but with GPUDirect disabled; (iv) \emph{UnpinnedMem-GPUDirect} is to copy via host unpinned memory with GPUDirect enabled; and (v) \emph{UnpinnedMem} is similar to (iv) but with GPUDirect disabled.  


\begin{figure*}[!htb]
\minipage{0.99\columnwidth}
\includegraphics[width=1.02\columnwidth]{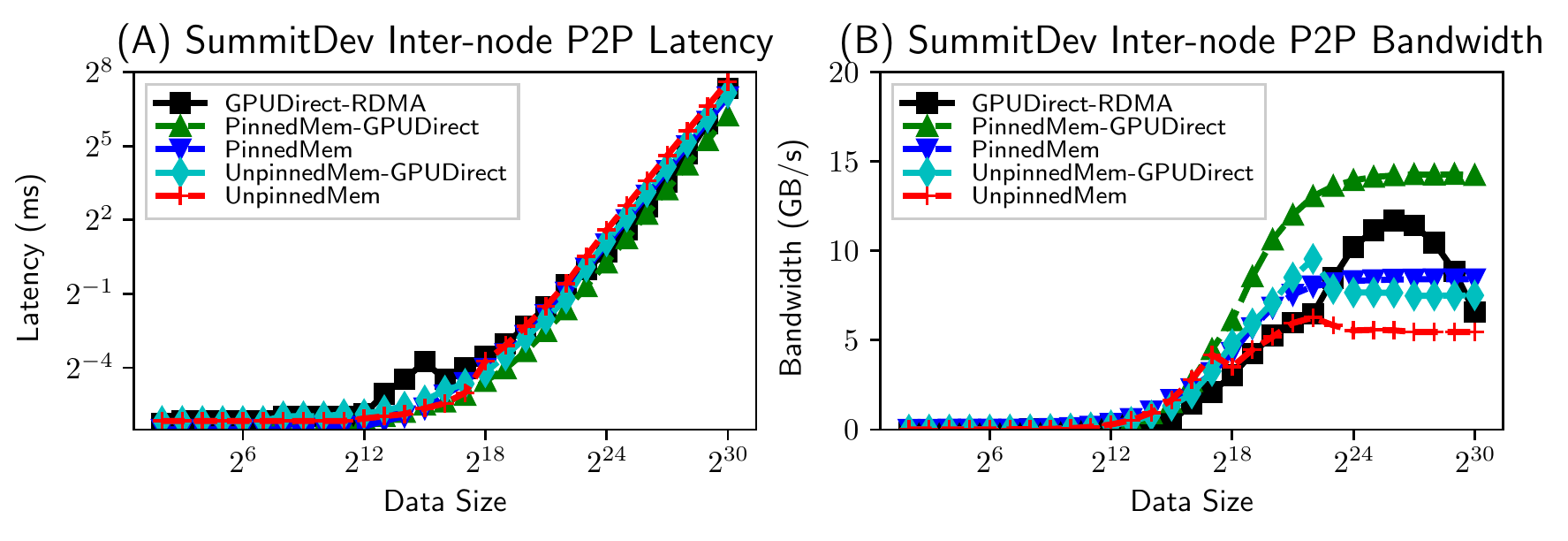} 
\caption{SummitDev inter-node P2P latency and bandwidth efficiency.}
\label{fig:ib-p2p}
\endminipage\hfill
\minipage{0.99\columnwidth}
\includegraphics[width=1.02\columnwidth]{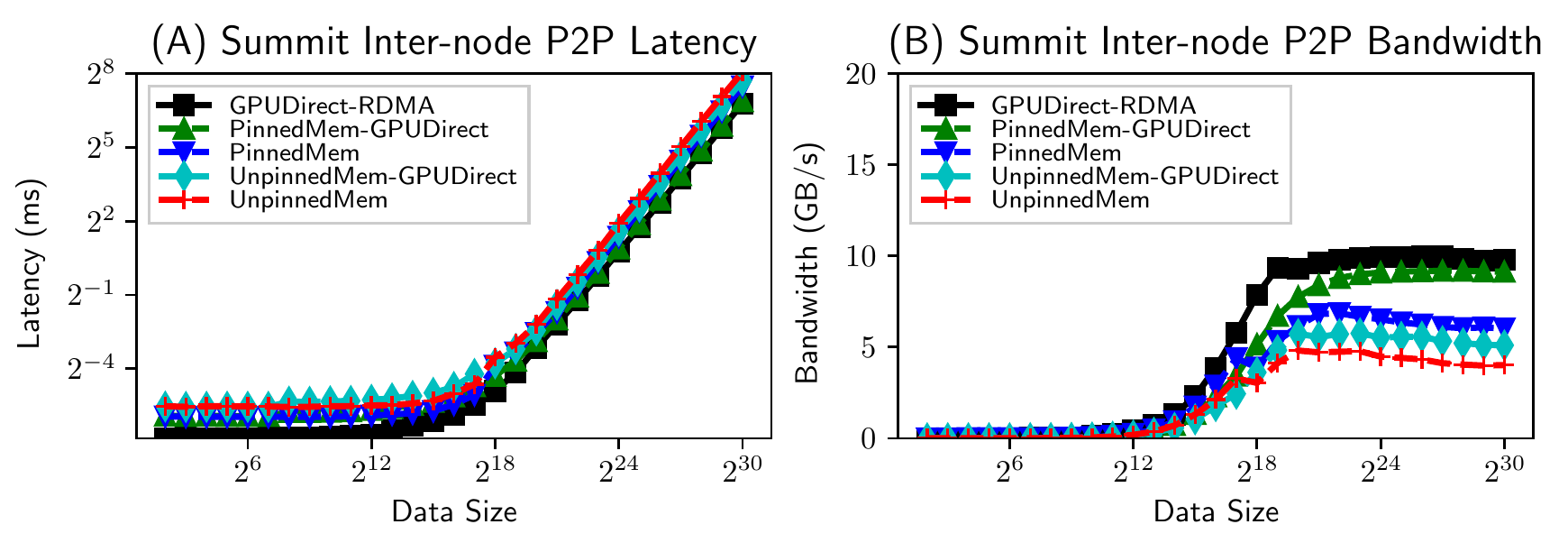} 
\caption{Summit inter-node P2P latency and bandwidth efficiency.}
\label{fig:ib-summit-p2p}
\endminipage
\end{figure*}

\begin{figure*}[!t]
\centering
\includegraphics[width=2.03\columnwidth]{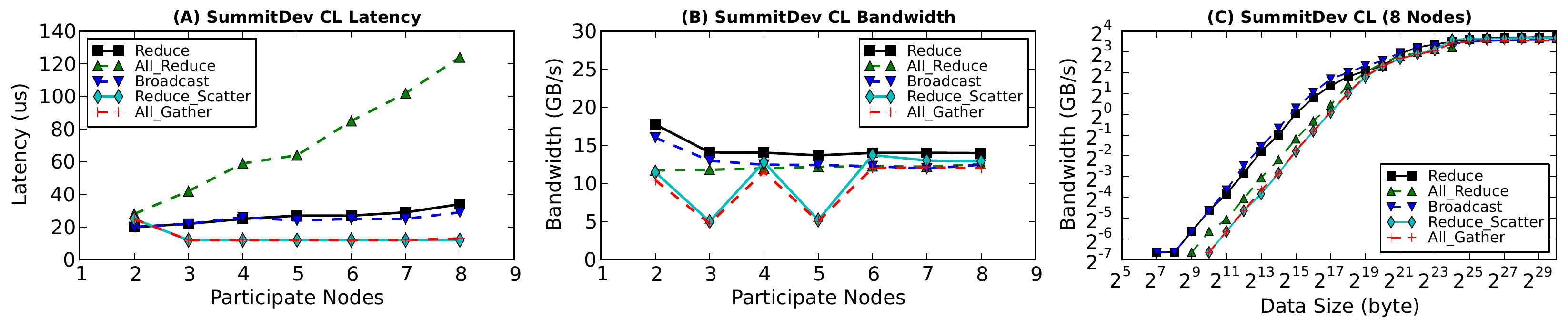} 
\caption{Inter-node CL communication latency and bandwidth with variable participant nodes, as well as bandwidth for 8 nodes with increased message size.}
\label{fig:ib-cl}
\end{figure*}

\begin{figure*}[!t]
\centering
\includegraphics[width=2.03\columnwidth]{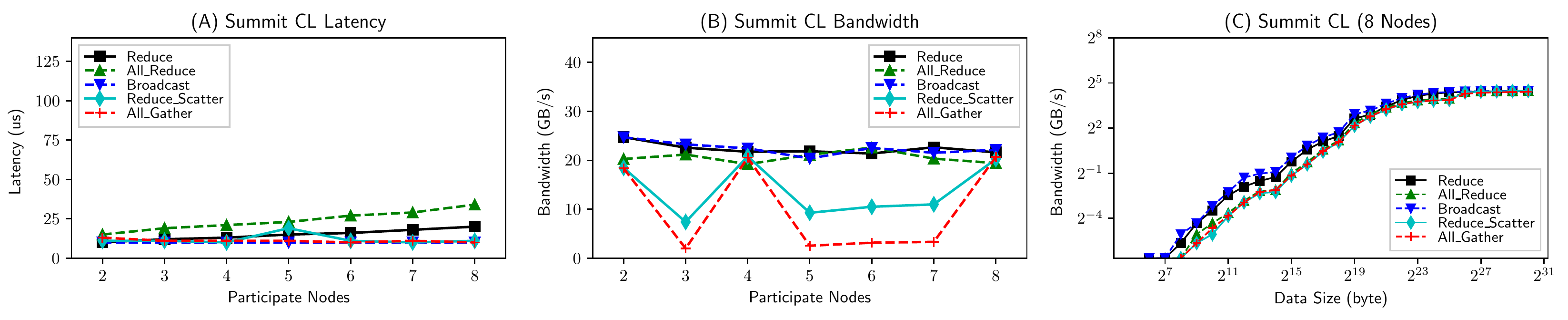} 
\caption{Inter-node CL communication latency and bandwidth with variable participant nodes, as well as bandwidth for 8 nodes with increased message size.}
\label{fig:ib-summit-cl}
\end{figure*}

The measured latency and bandwidth with respect to message size (from 4B to 1GB) under the five testing scenarios are illustrated in Figure~\ref{fig:ib-p2p} and \ref{fig:ib-summit-p2p}, for SummitDev and Summit, respectively. For better illustration, the latency curves use log-scale Y-axis while bandwidth curves use normal-scale Y. 

\vspace{4pt}\noindent \textbf{SummitDev:} From Figure~\ref{fig:ib-p2p}, we draw the following observations: (i) Until $2^{12}=4$KB, there is little difference among the five curves in terms of both latency and bandwidth; (ii) GPUDirect-RDMA shows the worst performance in the range from 4KB to 64KB for latency and from 4KB to 256KB for bandwidth, especially at 32KB. This is possibly due to limitations in some chipsets for P2P access through the CPU/IOH \cite{gpudirect-rdma}; (iii) From 4MB on, GPUDirect-RDMA shows its advantage on bandwidth and obtains its optimal bandwidth --- 12GB/s at 64MB. However, this is still lower than the PinnedMem-GPUDirect scheme, which demonstrates more than 14GB/s sustainable bandwidth with large message size; (v) it is also interesting to observe that the bandwidth of GPUDirect-RDMA actually degrades dramatically after 64MB, implying that breaking large messages into multiples of 64MB could be a better way to transfer in practice on SummitDev. 

\vspace{4pt}\noindent \textbf{Summit:} Regarding Figure~\ref{fig:ib-summit-p2p}, we have the following observations: (i) Unlike SummitDev, GPUDirect-RDMA shows the best performance among the five configurations on Summit: it always delivers the lowest latency, especially for small message size ($\le$1MB); it always exhibits the highest bandwidth, especially for large message size ($\ge$16KB). Meanwhile, the strange performance drop (i.e., latency increase and bandwidth degradation) in Figure~\ref{fig:ib-p2p} for SummitDev, disappear in Figure~\ref{fig:ib-summit-p2p} for Summit. These two points suggest that the technology of GPUDirect has been improved significantly from SummitDev to Summit, and becomes the best choice for GPU inter-node communication.


\subsection{Inter-Node Collective Communication}
\label{subsec_intercl}

Regarding inter-node collective communication, we measure the latency and bandwidth with respect to the number of participant nodes on both SummitDev and Summit. We tune the number of nodes from 2 to 8, with 1 GPU per node being utilized. Similarly, the startup latency is measured with 4B data transfer while the sustainable bandwidth is measured with sufficiently large data transfer (1GB). The latency results are shown in Figure~\ref{fig:ib-cl}-(A) and Figure~\ref{fig:ib-summit-cl}-(A) for SummitDev and Summit, respectively. The bandwidth results are shown in Figure~\ref{fig:ib-cl}-(B) and Figure~\ref{fig:ib-summit-cl}-(B). The bandwidth change with increasing message size is shown in Figure~\ref{fig:ib-cl}-(C) and Figure~\ref{fig:ib-summit-cl}-(C). We tried to illustrate the difference between enabling and disabling GPUDirect, but found that the results are in fact very similar. We suspect that GPUDirect-RDMA is internally enabled in NCCL-V2.

\vspace{4pt}\noindent \textbf{CL Latency:} As shown in Figure~\ref{fig:ib-cl}-(A), for SummitDev's IB fat-tree network, the latency-change for performing the five CL operations remains flat when scaling the number of nodes, except \emph{all-reduce}. Similar observation is also drawn in Figure~\ref{fig:ib-summit-cl}-(A) for Summit, the divergence is much less for \emph{all-reduce}. This may imply that it is a joint-effect of algorithm implementation in NCCL and the interconnect technology of GPUDirect. 

\vspace{4pt}\noindent \textbf{CL Bandwidth:} In terms of bandwidth in Figure~\ref{fig:ib-cl}-(B), similar to NVLink (Figure~\ref{fig:nccl_bandwidth}-(B) and (D)), strong NUMA effects emerge under 3 and 5 nodes for \emph{reduce\_scatter} and \emph{all\_gather} on SummitDev. And for Summit, this happens under 3, 5, 6, 7 nodes in Figure~\ref{fig:ib-summit-cl}. Nevertheless, the bandwidth overall remains unchanged. This is different from the bandwidth scenarios exhibited by both PCIe (decreasing) and NVLink (increasing) in the inter-node P2P communication study. 

\vspace{4pt}\noindent \textbf{CL Efficiency on Message Size:} Finally, the bandwidth for the five CL operations converge and saturate around 32/64MB message size, demonstrating nearly 16/32 GB/s sustainable peak bandwidth on SummitDev in Figure~\ref{fig:ib-summit-cl}-(C), and Summit in Figure~\ref{fig:ib-summit-cl}-(C), respectively. Overall, Summit delivers much better GPU inter-node communication performance than SummitDev.

%
%
%

\begin{table*}[!t] 
\centering\scriptsize
\caption{Tartan Benchmark Suite.} 
\begin{tabular}{|c|l|c|c|c|c|c|} \hline 
\textbf{App} & \textbf{Brief Description} & \textbf{abbr.} & \textbf{Domain} &  \textbf{Comm} & \textbf{Scaling} & \textbf{Pattern}  \\ \hline

\emph{ConvNet2} & Convolution neural networks via data, model and hybrid parallelism &  \texttt{CNN} & \emph{Machine Learning} & CUDA & Scale-up & P2P  \\ \hline
\emph{Cusimann} & Global optimization via parallel simulated annealing algorithm &  \texttt{CSM} & \emph{Optimization} & OpenMP & Scale-up & CPU-GPU  \\ \hline
\emph{GMM} & Multivariate data clustering via Expectation Maximization with Gaussian mixture model &  \texttt{GMM} & \emph{Data Analysis} & OpenMP & Scale-up & CL-Broadcast   \\ \hline
\emph{Kmeans} & Kmeans clustering for double-precision data on multi-GPUs attached to the same node &  \texttt{KMN} & \emph{Data Analysis} & CUDA & Scale-up & CL-AllReduce  \\ \hline
\emph{MonteCarlo} & Monte Carlo option pricing from CUDA SDK &  \texttt{MTC} & \emph{Finance} & CUDA & Scale-up & CPU-GPU   \\ \hline
\emph{Planar} & Depth-first-search (DFS) and backtracking to solve Planar Langford's Sequences &  \texttt{PLN} & \emph{Number Theory} & CUDA & Scale-up & CPU-GPU  \\ \hline
\emph{Trueke} & Exchange Monte Carlo for 3D random field Ising model&  \texttt{TRK} & \emph{HPC Simulation} & OpenMP & Scale-up & CL-Broadcast   \\ \hline\hline

\emph{B2rEqwp} & 3D earthquake wave-propogation model simulation using 4-order finite difference method
 &  \texttt{BRQ} & \emph{HPC Simulation} & MPI & Scale-out & P2P   \\ \hline
\emph{Diffusion} & A multi-GPU implementation of 3D Heat Equation and inviscid Burgers' Equation &  \texttt{DFF} & \emph{HPC Simulation} & MPI & Scale-out & P2P   \\ \hline
\emph{Lulesh} & Livermore unstructured Lagrangian explicit shock hydrodynamics &  \texttt{LLH} & \emph{Molecular Dynamics} & MPI & Scale-out & P2P   \\ \hline
\emph{CoMD} & A reference implementation of classical molecular dynamics algorithms and workloads &  \texttt{CMD} & \emph{Molecular Dynamics} & MPI & Scale-out & P2P/CL   \\ \hline
\emph{Prbench} & Page rank computation by multi-GPUs &  \texttt{PRB} & \emph{Graph Processing} & MPI & Scale-out & P2P/CL  \\ \hline
\emph{HIT} & Simulating Homogeneous Isotropic Turbulence by solving Navier-Stokes equations in 3D &  \texttt{HIT} & \emph{HPC Simulation} & MPI & Scale-out & CL All-to-All  \\ \hline
\emph{Matvec} & Matrix multiplication via mpi-scatter, broadcast and gather &  \texttt{MAM} & \emph{Linear Algebra} & MPI & Scale-out & CL  \\ \hline

\end{tabular} 
\label{tab:bench} 
\end{table*} 

\section{GPU Interconnect Benchmarking}

The microbenchmarking exhibit some basic characteristics of modern GPU interconnects. However, in terms of real multi-GPU applications, their impact remains unknown. In this section, we use the Tartan Benchmark Suite \cite{li2018tartan} to evaluate the impact of the GPU interconnect. The applications are listed in Table~\ref{tab:bench}. Particularly, we focus on two aspects: (1) the impact of a faster GPU interconnect such as NVLink, compared with PCIe on intra-node scale-up applications; (2) the impact of GPUDirect-RDMA on inter-node scale-out applications. We perform two types of scaling measurement: (a) \emph{strong scaling}, which fixes the problem size and measures the time reduction when increasing the number of GPUs; (b) \emph{weak scaling}, which measures the time reduction when increasing the number of GPUs with fixed problem size per GPU. For overall performance, we use the entire application speedup (measured by CPU-side \texttt{time} command for whole application elapsed-time) as the performance metric, making a fair comparison across applications. For scale-up applications, we use the vendor-provided \texttt{nvprof} to measure the three types of GPU communication: \emph{HostToDevice (H2D)}, \emph{DeviceToHost (D2H)} and \emph{DeviceToDevice (D2D)} in order to gain more knowledge about the underlying communication pattern. All the reported data points of the figures in this section are the average results of multiple times' execution.



\subsection{Intra-node Scale-up}

\begin{figure*}[!t]
\centering
\includegraphics[width=2\columnwidth]{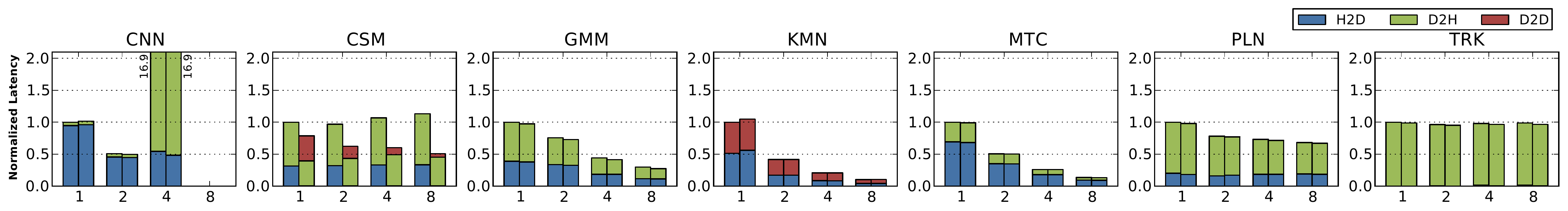} 
\caption{Normalized latency reduction by NVLink-V1 and NCCL-V2 of strong scaling for single-node scaling-up on NVIDIA P100-DGX-1.}
\label{fig:dgx1_scale_up_strong}
\end{figure*}

\begin{figure*}[!t]
\centering
\includegraphics[width=2\columnwidth]{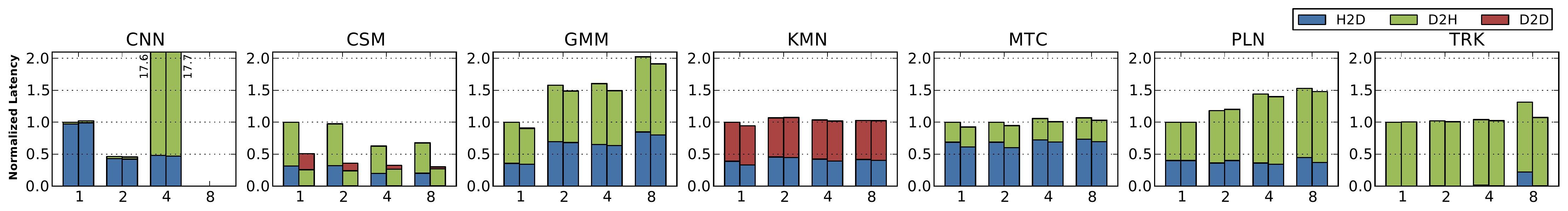} 
\caption{Normalized latency reduction by NVLink-V1 and NCCL-V2 of weak scaling for single-node scaling-up on NVIDIA P100-DGX-1.}
\label{fig:dgx1_scale_up_weak}
\end{figure*}

\begin{figure*}[!t]
\centering
\includegraphics[width=2\columnwidth]{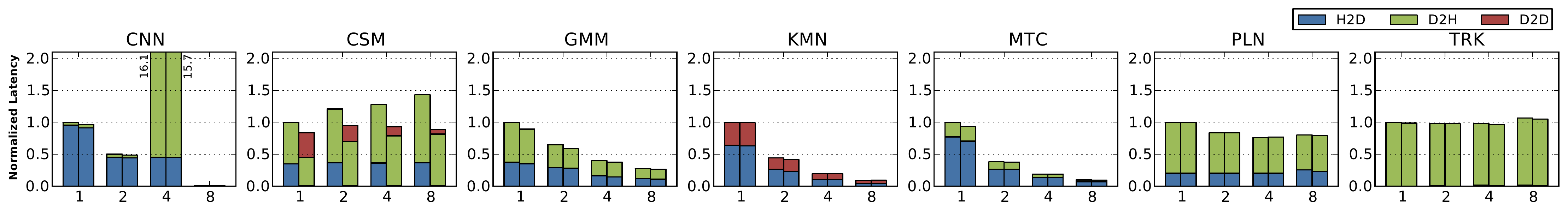} 
\caption{Normalized latency reduction by NVLink-V2 and NCCL-V2 of strong scaling for single-node scaling-up on NVIDIA V100-DGX-1.}
\label{fig:dgx2_scale_up_strong}
\end{figure*}

\begin{figure*}[!t]
\centering
\includegraphics[width=2\columnwidth]{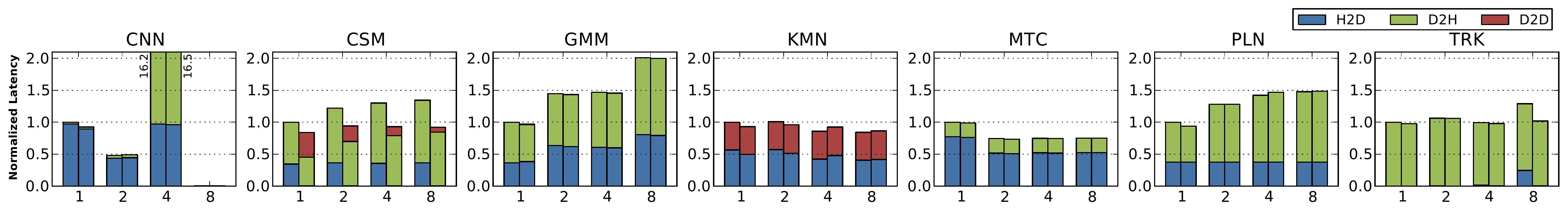} 
\caption{Normalized latency reduction by NVLink-V2 and NCCL-V2 of weak scaling for single-node scaling-up on NVIDIA V100-DGX-1.}
\label{fig:dgx2_scale_up_weak}
\end{figure*}

For intra-node scale-up scenarios, we evaluated the seven scale-up applications on P100-DGX-1 and V100-DGX-1, with and without NVLinks. Since many of these applications are hard-coded to leverage all the available GPUs in the system, we configure the system environment through \emph{export CUDA\_VISIBLE\_DEVICES=x} to manipulate the number of GPUs being visible to the applications. Figure~\ref{fig:dgx1_scale_up_strong} and \ref{fig:dgx1_scale_up_weak} illustrate the break out of the latency for the three types of communication (i.e., H2D, D2H, and D2D) regarding the original implementation (i.e., \emph{Baseline}) and our modification (i.e., via \emph{NCCL}) as described in Section~IV-A, for intra-node strong and weak scaling on P100-DGX-1. The intention is that the original implementation will show the performance of PCIe, while our modification would convert a big portion of CPU-GPU communication to GPU-GPU communication, so as to show the performance gain from NVLink. Figure~\ref{fig:dgx2_scale_up_strong} and \ref{fig:dgx2_scale_up_weak} show the results on V100-DGX-1. The performance change of the entire application is given in the supplementary file due to space limitation.

\vspace{3pt}\noindent\textbf{Impact of NVLink.} Although our observation from microbenchmarking in Section~\ref{sec_evaluation} show that NVLink can significantly improve inter-GPU communication efficiency, based on these figures, it is clear that those improvements do not directly transform into overall communication latency reduction, nor the whole application speedup (see the supplementary file); except \texttt{CSM} and \texttt{GMM}, there is not very significant difference between the Baseline and NCCL bars for both platforms. There are several reasons behind this. First, based on our experience on assessing the over 50 multi-GPU application candidates, most of those scale-up cases are based on master-slave programming model, where the CPU is the master, handling the sequential portions of code and GPUs are the slaves, processing the parallel portions. Under this model, communication only occurs between CPU and GPUs; no inter-GPU transaction is presumed. In addition, the CPU-GPU communication is also highly optimized. This is true for \texttt{CSM}, \texttt{GMM}, \texttt{KMN}, \texttt{MTC}, \texttt{PLN} and \texttt{TRK}. For \texttt{CSM}, \texttt{PLN}, \texttt{GMM}, and \texttt{TRK}, we manage to convert some CPU-GPU communication into GPU-GPU communication through NCCL. As can be seen, for \texttt{CSM}, the effect is obvious: a large portion of the D2H and H2D communication is replaced by D2D. For \texttt{GMM}, although we gained $\sim$6\% latency reduction, from the \emph{nvprof} trace file, we did not observe any D2D communication transactions (but rather D2H and H2D) before/after NCCL's \emph{BroadcastKernelSmall()} kernels. This is potentially caused by the flexible design strategy adopted by NCCL-V2 to efficiently leverage all available interconnect bandwidth (Section~III-B). When the data size per transmission is small, it may not choose D2D communication via NVLink. Similar conditions also observed for \texttt{PLN} and \texttt{TRK}. For \texttt{KMN} and \texttt{MTC}, there is no GPU-GPU communication at all. The D2D in \texttt{KMN} is actually local data movement within the same GPU. \texttt{CNN} is another case. For \texttt{CNN}, the figures do not show bars under 8 GPUs because \texttt{CNN} implementation requires arbitrary data copying among arbitrary peers at runtime, which currently fails when 8 GPUs are utilized, since not every two GPUs are directly connected by NVLink under the current NVLink topology. As it internally uses \emph{cudaMemcpyDefault} and \emph{unified-space} for data copying across GPUs, NVLink is already internally enabled by CUDA for an NVLink-equipped machine such as DGX-1. We tried to modify the \texttt{CNN} code so that PCIe can be essentially enforced but did not run correctly with success. This is why the two bars of \texttt{CNN} exhibit the same value for all four figures. It is also interesting to see that when more than 4 GPUs participant in the computation, the D2H communication increases dramatically, potentially due to the gradient merging overhead in the backpropagation. Secondly, since today's scale-up applications are mostly based on the master-slave programming model, communication often only accounts for a small fraction of the total execution time, let alone the inter-GPU communication which tended to be avoided previously when creating applications, thus hardly become the system bottleneck. Finally, employing NVLink (either P2P or CL) introduces additional overhead (e.g., enable/disable peer access, routing, NCCL initialization, etc).

To summarize, a faster GPU interconnect such as NVLink has been reported to be beneficial for accelerating modern deep-learning frameworks \cite{shams2017evaluation, tallent2017evaluating}. However, regarding general GPGPU applications, without (i) replacing the underlying CPU-centric master-slave programming model by a more distributed parallelization model, or (ii) migrating the communication master role to a GPU (e.g., off-loading GPU communication control from CPU to GPU via techniques such as \emph{NVSHMEM} \cite{potluri2017gpu}), optimized inter-GPU communication via faster intra-node GPU interconnect such as NVLinks can hardly become significant enough to lift the entire application's speedup. Therefore, we believe that this observation paves the road for developing interconnect-friendly programming models for multi-GPU scale-up scenarios so that faster interconnect (e.g., NVLinks) can truly play a role in improving the overall application efficiency.

\subsection{Inter-node Scale-out}

\begin{figure*}[!t]
\centering
\includegraphics[width=2.05\columnwidth]{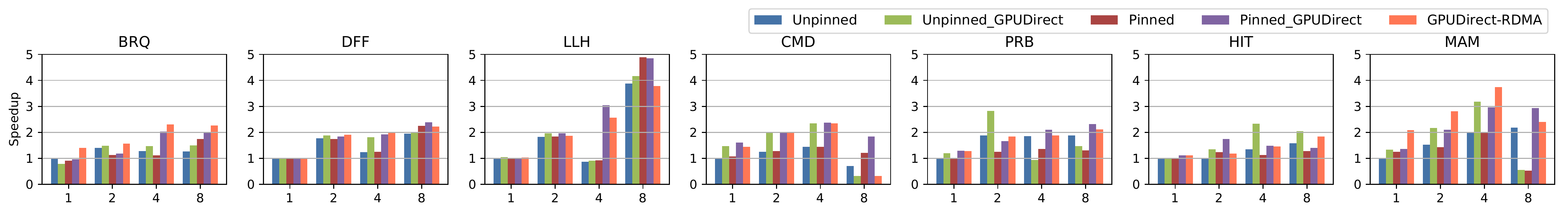} 
\caption{Performance speedup by InfiniBand GPUDirect-RDMA of strong scaling for multi-node scaling-out on ORNL SummitDev.}
\label{fig:summitdev_strong}
\end{figure*}

\begin{figure*}[!t]
\centering
\includegraphics[width=2.05\columnwidth]{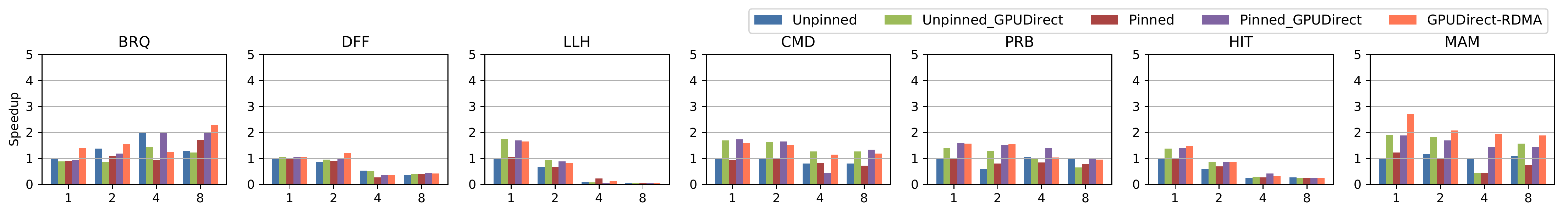} 
\caption{Performance speedup by InfiniBand GPUDirect-RDMA of weak scaling for multi-node scaling-out on ORNL SummitDev.}
\label{fig:summitdev_weak}
\end{figure*}

\begin{figure*}[!t]
\centering
\includegraphics[width=2.05\columnwidth]{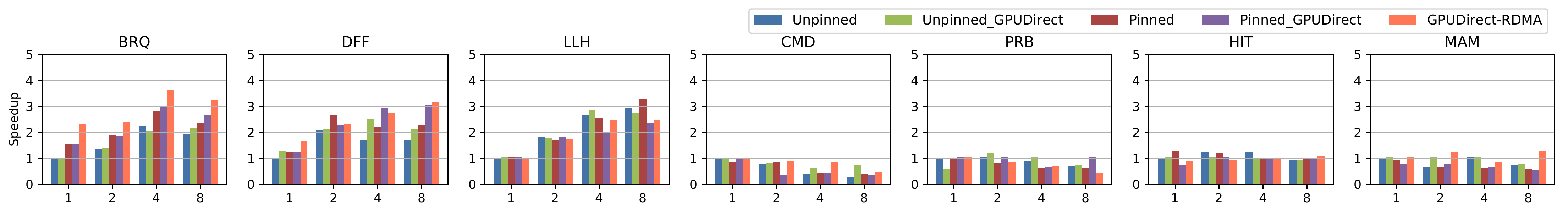} 
\caption{Performance speedup by InfiniBand GPUDirect-RDMA of strong scaling for multi-node scaling-out on ORNL Summit.}
\label{fig:summit_strong}
\end{figure*}

\begin{figure*}[!t]
\centering
\includegraphics[width=2.05\columnwidth]{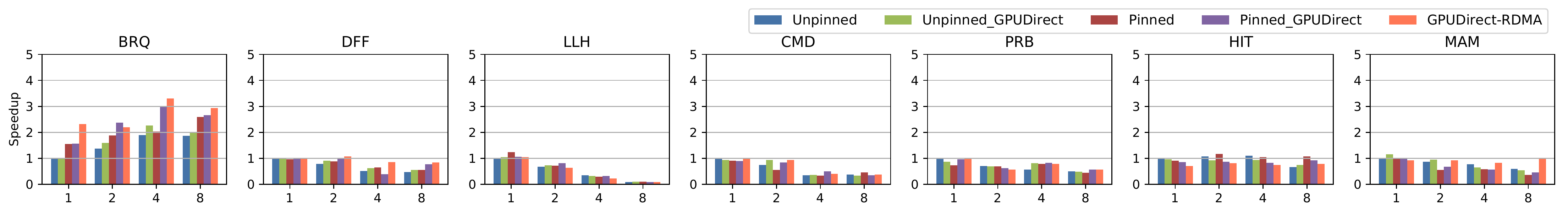} 
\caption{Performance speedup by InfiniBand GPUDirect-RDMA of weak scaling for multi-node scaling-out on ORNL Summit.}
\label{fig:summit_weak}
\end{figure*}


For inter-node scale-out scenarios, we run the seven scale-out applications from Tartan on \emph{SummitDev} and \emph{Summit} (see Table~\ref{tab:platform}), with each MPI rank binding to a node using only a single GPU. Similar to the discussion in Section~\ref{subsec_interp2p}, we measured the overall application performance under five scenarios: \emph{GPUDirect-RDMA},  \emph{PinnedMem-GPUDirect},  \emph{PinnedMem}, \emph{UnpinnedMem-GPUDirect} and \emph{UnpinnedMem}.

Figure~\ref{fig:summitdev_strong} and \ref{fig:summitdev_weak} illustrate the speedups with respect to single-node \emph{UnpinnedMem} for strong and weak scaling tests, respectively, on SummitDev. Figure~\ref{fig:summit_strong} and \ref{fig:summit_weak} illustrate the speedups for strong and weak scaling on Summit. As can be seen, compared with the intra-node scale-up cases, the MPI-based inter-node scale-out applications exhibit much better scaling behavior in both strong and weak scaling tests, implying that compared with the intra-node fast interconnect, the inter-node network is much easier to become the system bottleneck. Improving inter-node network speed can lead to significant performance gain for multi-GPU applications. 

Regarding to GPUDirect-supported IB interconnect, we have the following observations: (i) Enabling GPUDirect can bring immediate performance enhancement, whether or not the transmitted data reside in CPU memory or GPU memory; (ii) Using pinned memory is also beneficial, especially in coordination with GPUDirect enabled; (iii) GPUDirect+RDMA can be especially helpful in certain applications (e.g., \texttt{BRQ} and \texttt{MAM}) for SummitDev, and overall for Summit. This is potentially due to the significant improvement on GPUDirect-RDMA performance in Summit than SummitDev; (iv) Overall, the multi-node performance scaling on Summit is less obvious than on SummitDev, especially for strong scaling. This is not due to any degradation in communication efficiency in Summit, but essentially the reverse: Summit improves the fundamental communication bandwidth, which enhances the baseline performance. That is why the relative speedups drop on Summit, compared to SummitDev. Overall, we suggest the application developers to adopt \emph{PinnedMem-GPUDirect} for SummitDev, and \emph{GPUDirect-RDMA} for Summit.

All in all, for scale-out applications to benefit from a faster inter-node interconnect (e.g., IB-RDMA), the major difficulty is not from the hardware or the application, but from the communication abstract interfaces such as MPI. If a new MPI implementation can internally integrate NCCL, further harvesting multi-GPU interconnect performance (e.g., NVLink and IB-RDMA) can be much more easier. Again, initiating communication completely on the GPU side without CPU intervention (e.g., via \emph{GPUDirect-Async} \cite{agostini2018gpudirect} or \emph{GPU-triggered networking} \cite{lebeane2017gpu}) may also be critical for good GPU performance delivery.


%
\section{Discussion}
\label{sec_dis}

Modern GPU interconnect technologies such as NVLink are claimed to be transparent but in reality it is more complicated to be leveraged for high performance. (1) \emph{NUMA effect}. Among the five types of intra-node GPU interconnect techniques, PCIe, NVLink-V1 and V2 show strong NUMA effect in the tested platforms, due to various reasons including topology, position, connectivity, routing, sharing, chipset, etc. NVSwitch and NV-SLI show UMA. (2) \emph{Heterogeneity}. All the tested platforms incorporate more than one type of interconnect network. These networks have their own characteristics and can work exclusively, concurrently, or cooperatively, depending on the system design. Therefore, one has to handle the interconnect heterogeneity: choosing one interconnect over the others (e.g., NVLink shows strong advantage on bandwidth rather than latency over PCIe), leveraging them simultaneously, or cooperatively integrate them as an inclusive solution. This is especially difficult at runtime. (3) \emph{Communication Efficiency}. There are several factors restricting the delivery of optimal communication performance, including message size, system design (dual isolated subnetworks in Summit and SummitDev), hardware limitation (e.g., PCIe antilocality, GPUDirect-RDMA in SummitDev), and library implementation (e.g., NCCL on DGX-1 with 3 and 5 nodes). All these lead to the difficulties in leveraging modern GPU interconnect for high-efficient inter-GPU communication.
 
Our evaluation motivates the following research directions: (1) \emph{Developing novel multi-GPU programming models}. Existing multi-GPU programming models rely on CPU-oriented parallel programming models, such as OpenMP and MPI, to manage multiple GPUs. Consequently, either there is a mismatch (e.g., CPU-master-GPU-slave model can hardly benefit from inter-GPU network), or there is a legacy in adopting new GPU interconnect technologies (e.g., integrating NCCL into MPI, as it is shown that NCCL delivers better communication efficiency than MPI on all-reduce \cite{yin2018}). Therefore, new multi-GPU programming models are highly desired, especially those that are adaptive, portable, tractable, and being able to effectively address the complexity aforementioned. In addition, existing multi-GPU algorithms are usually designed to minimize or even eliminate communications, due to the huge performance gap between local access and remote access (e.g., via PCIe). With the emergence of new GPU interconnect, and foreseeing their fast development trend, it may be the right time to reconsider the role of inter-GPU communication when designing new parallel models and algorithms. (2) \emph{Developing practical multi-GPU performance models.} This is for performance prediction, optimization, and analytics in multi-GPU application development and tuning. In addition, an appropriate performance model is crucial for GPU task allocation, scheduling and migration in a shared environment (e.g., Cloud and HPC centers). (3) Developing new communication patterns and libraries for better matching the underlying interconnect and delivering high-performance. For example, regarding the dual-subnetwork interconnect topology for NVLink in Summit, it is worth to figure out how to efficiently distribute and exchange data among the two islands, taking data reuse and X-bus bandwidth into consideration. Give another example, a recent work shows that a 2D-Torus communication pattern can deliver better communication efficiency than NCCL's ring pattern for all-reduce. This new pattern can be obviously migrated from inter-node to the intra-node NVLink interconnect in DGX-1s, or NVSwitch in DGX-2, where multiple communication ring-paths can be constructed. This is especially desired when porting traditional CPU-based HPC applications onto the new GPU-based exascale systems, such as Summit \cite{summit}, Sierra \cite{sierra} and Perlmutter \cite{perlmutter}. As part of the community effort, we are planning to pursue these research directions in our future work with our past experience on GPU analytic modeling \cite{li2015transit, li2016x, li2017exploring} and performance optimization \cite{li2015adaptive, li2015fine, li2016sfu, li2016critical, liu2016synchronization, li2016optimizing, li2017locality, li2017bvf, li2018warp, shen2018cudaadvisor, wang2018superneurons}.

\section{Related Work}
\label{sec_relatedwork}

\noindent\textbf{Intra-node GPU Computing}.  Spafford et al. \cite{spafford2011quantifying} analyzed the NUMA effects in a multi-GPU node and provided optimization guidance. Kim et al. \cite{kim2014multi} proposed to rely on hybrid-memory-cubes (HMCs) to build a memory network for simplifying multi-GPU memory management and improving programmability. Wang et al. \cite{wang2014gpu} presented a design to realize GPU-Aware MPI to support data communication among intra-node GPUs with standard MPI. Ben-Nun et al. \cite{ben2015memory} described an automatic multi-GPU partition framework to distribute workload based on their memory access patterns. Cabezas et al. \cite{cabezas2015automatic} showed a software solution, including programming interfaces, compiler support and runtime, to partition GPU kernels for multi-GPU execution in a single node. Finally, Sun et al. \cite{sun2018evaluating} evaluated the potential performance benefit and tradeoffs of AMD's \emph{Radeon Open Compute} (ROC) platform for \emph{Heterogeneous System Architecture} (HSA).

\vspace{3pt}\noindent \textbf{Multi-node GPU Computing}. For MPI-based multi-node GPU computing, Wang et al. \cite{wang2011mvapich2} introduced a MPI design that integrates CUDA data movement transparently with MPI. Gysi et al. \cite{gysi2016dcuda} proposed a hardware approach to overlap computation and communication in a GPU cluster. Klenk et al. \cite{klenk2017relaxations} analyzed the exascale proxy applications on their communication patterns and proposed a matching algorithm for GPUs to comply with MPI constraints. Awan et al. \cite{awan2017optimized} proposed a pipelined chain design for MPI broadcast collective operations on multi-GPU nodes to facilitate various deep learning frameworks. 



\section{Conclusion}
\label{sec_conclusion}

In this paper, we characterize and evaluate six types of modern GPU interconnects, including PCIe, NVLink-V1, NVLink-V2, NV-SLI, NVSwitch, and InfiniBand with GPUDirect-RDMA, using the Tartan Benchmark Suite over six GPU servers and HPC platforms: NVIDIA's P100-DGX-1, V100-DGX-1, DGX-2, RTX2080-SLI systems, and ORNL's SummitDev and Summit supercomputers, covering both Peer-to-Peer and Collective communication patterns. We addressed four new types of NUMA effects for intra-node GPU communication, and proposed some insightful observations for enabling practical optimization guidelines. This evaluation study attempts to help the HPC community to push forward multi-GPU research and development, particularly the construction of more mature multi-GPU programming, execution, and performance models.


\ifCLASSOPTIONcompsoc
  \section*{Acknowledgments}
\else
  \section*{Acknowledgment}
\fi

This research was supported by the Application Assessment program within the Exascale Computing Project (17-SC-20-SC), a collaborative effort of the U.S. Department of Energy Office of Science and the National Nuclear Security Administration. This research was supported by the U.S. DOE Office of Science, Office of Advanced Scientific Computing Research, under award 66150: ``CENATE - Center for Advanced Architecture Evaluation''. This research was supported by the High Performance Data Analytics (HPDA) program at PNNL. This research used resources of the Oak Ridge Leadership Computing Facility at the Oak Ridge National Laboratory, which is supported by the Office of Science of the U.S. Department of Energy under Contract No. DE-AC05-00OR22725. The Pacific Northwest National Laboratory is operated by Battelle for the U.S. Department of Energy under contract DE-AC05-76RL01830.

\ifCLASSOPTIONcaptionsoff
  \newpage
\fi

\bibliographystyle{IEEEtran}
\bibliography{eva}

%
%







\end{document}